\documentclass[12pt,preprint]{emulateapj}

\slugcomment{Astronomical Journal}
\shorttitle{HCN/HCO+ in four LIRGs}
\shortauthors{Imanishi et al.}

\begin{document}

\title{Nobeyama Millimeter Interferometric HCN(1--0) and HCO$^{+}$(1--0)
Observations of Further Luminous Infrared Galaxies} 

\author{Masatoshi Imanishi \altaffilmark{1}}
\affil{National Astronomical Observatory, 2-21-1, Osawa, Mitaka, Tokyo
181-8588, Japan} 
\email{masa.imanishi@nao.ac.jp} 

\author{Kouichiro Nakanishi \altaffilmark{1}}
\affil{Nobeyama Radio Observatory, Minamimaki, Minamisaku, Nagano,
384-1305, Japan} 

\author{Yoichi Tamura \altaffilmark{2}}
\affil{Department of Astronomy, The University of Tokyo, 7-3-1
Hongo, Bunkyo-ku, Tokyo 113-0033, Japan} 

\and

\author{Chih-Han Peng}
\affil{formerly Department of Astronomy, School of Science, Graduate
University for Advanced Studies, Mitaka, Tokyo 181-8588} 

\altaffiltext{1}{Department of Astronomy, School of Science, Graduate
University for Advanced Studies, Mitaka, Tokyo 181-8588}

\altaffiltext{2}{National Astronomical Observatory of Japan, 2-21-1
Osawa, Mitaka, Tokyo 181-8588, Japan} 

\begin{abstract}

We report the results of interferometric HCN(1--0) and
HCO$^{+}$(1--0) observations of four luminous infrared galaxies (LIRGs),
NGC 2623, Mrk 266, Arp 193, and NGC 1377, as a final sample of our
systematic survey using the Nobeyama Millimeter Array. 
Our survey contains the most systematic {\it interferometric,
spatially-resolved, simultaneous} HCN(1--0) and HCO$^{+}$(1--0)
observations of LIRGs.  
Ground-based infrared spectra of these LIRGs are also presented to
elucidate the nature of the energy sources at the nuclei. 
We derive the HCN(1--0)/HCO$^{+}$(1--0) brightness-temperature ratios of
these LIRGs and confirm the previously discovered trend that LIRG nuclei
with luminous buried AGN signatures in infrared spectra tend to show
high HCN(1--0)/HCO$^{+}$(1--0) brightness-temperature ratios, as seen in
AGNs, while starburst-classified LIRG nuclei in infrared spectra
display small ratios, as observed in starburst-dominated galaxies.  
Our new results further support the argument that the
HCN(1--0)/HCO$^{+}$(1--0) brightness-temperature ratio can be used to
observationally separate AGN-important and starburst-dominant galaxy
nuclei.
\end{abstract}

\keywords{galaxies: active --- galaxies: nuclei ---  galaxies: ISM ---
radio lines: galaxies --- galaxies: individual (\objectname{NGC 2623,
Mrk 266, Arp 193, and NGC 1377})} 

\section{Introduction}

Luminous infrared galaxies (LIRGs) radiate the bulk of their large
luminosities ($L> 10^{11}L_{\odot}$) as infrared dust emission.
Large infrared luminosities mean that: (1) luminous energy
sources are present, but are hidden behind dust; (2) energetic radiation
from the hidden energy sources is absorbed by the surrounding dust; and
(3) the heated dust grains re-emit this energy as infrared thermal
radiation. To understand the nature of LIRGs, it is essential to unveil
the hidden energy sources; namely, to distinguish whether starbursts (=
nuclear fusion inside stars) are dominant, or active galactic nuclei
(AGNs; active mass accretion onto a central compact supermassive black
hole [SMBH] with $>$10$^{6}$M$_{\odot}$) are also energetically important.  

Unlike AGNs surrounded by toroidally shaped dust, which are classified
optically as Seyferts \citep{vei87}, 
concentrated molecular gas and dust in LIRG nuclei \citep{sam96} can easily
{\it bury} (= obscure in virtually all directions) the putative compact
AGNs, making AGN signatures very difficult to detect optically. 
However, it is crucial to understand the energetic role of such
optically elusive {\it buried} AGNs in LIRG nuclei. 

A starburst (nuclear fusion) and buried AGN (mass accretion
onto a SMBH) have very different energy generation mechanisms.
Specifically, while UV emission is predominant in
a starburst, an AGN emits strong X-ray emission in addition to UV.
Additionally, in a normal starburst, the stellar energy sources and dust
are spatially well mixed, so energy sources are spatially extended. 
Since the energy generation efficiency of the nuclear fusion reaction is
only $\sim$0.5\% of Mc$^{2}$ (M is the mass of material used in the nuclear
fusion reaction), the emission surface brightness of a 
starburst is modest and has both observational \citep{soi00} and 
theoretical \citep{tho05} upper limits ($\sim$10$^{13}$L$_{\odot}$ 
kpc$^{-2}$). 
On the other hand, in an AGN, the mass accreting SMBH is spatially very
compact, and thus more centrally concentrated than the surrounding gas
and dust. 
The high energy generation efficiency of an AGN (6--42\% of Mc$^{2}$; 
M is the mass of accreting material; Thorne 1974) 
can generate large luminosities from a very compact region, producing 
a very high emission surface brightness \citep{soi00}.
These differences between an AGN and a starburst could create differences 
in the properties of the surrounding molecular gas and dust that may
be distinguishable based on observations at wavelengths of low dust
extinction.

Molecular gas emission lines in the millimeter wavelength 
range can be used effectively to investigate the hidden energy sources
of LIRG nuclei. 
First, dust extinction is very small. 
Second, theoretical calculations predict that X-ray- and
UV-emitting energy sources show different behaviors of 
molecular line emission due to different chemical reactions
\citep{mei06}. 
Finally, mid-infrared 10--20 $\mu$m emission is strong in an AGN because
of hot dust thermal emission heated by the high surface brightness
energy source of an AGN. 
The strong mid-infrared emission could selectively enhance particular 
molecular line emission through an infrared radiative pumping mechanism
\citep{aal95,gar06,gue07,wei07,aal07}. 
In this paper, we focus on HCN(1--0) and
HCO$^{+}$(1--0) lines, because an AGN could enhance HCN(1--0) emission
through an increased HCN abundance \citep{lin06} and/or infrared
radiative pumping \citep{aal95,gar06,wei07}. 
In fact, \citet{koh05} found observationally that HCN(1--0) emission is
stronger, relative to HCO$^{+}$(1--0), in AGN-dominated nuclei than in 
starburst galaxies, demonstrating the potential of this HCN(1--0) and 
HCO$^{+}$(1--0) based method for the purpose of distinguishing between
an AGN and a starburst. 
This method may be effective even for a Compton thick (N$_{\rm H}$
$>$ 10$^{24}$ cm$^{-2}$) buried AGN, the detection of which is very difficult 
with direct X-ray observations. 

HCN(1--0) and HCO$^{+}$(1--0) observations of a large number of LIRGs 
have been made, using single-dish radio telescopes \citep{gao04,gra08,kri08}.  
However, the beam sizes of these observations are so large ($>$25 arcsec) that
severe contamination from spatially-extended star-forming emission is
unavoidable, possibly hampering the detection of the AGN signatures of
LIRG nuclei, where putative AGNs are expected to be present.
Furthermore, LIRGs often show multiple-nucleus morphologies with small
separations ($<$10 arcsec), for which spatially-resolved study is not
possible with single-dish radio telescopes. 
Finally, the HCN(1--0) and HCO$^{+}$(1--0) data were collected at different
times under different weather conditions, possibly increasing
uncertainty regarding their relative strengths due to the ambiguity of
inter-calibration of data.  

To overcome these issues, we performed millimeter {\it interferometric, 
spatially-resolved, simultaneous} HCN(1--0) and HCO$^{+}$(1--0)
observations of LIRGs, using the Nobeyama Millimeter Array (NMA) 
\citep{ima04,ink06,in06,ima07b}. 
In our study, only nuclear HCN(1--0) and HCO$^{+}$(1--0) emission
was extracted to investigate their ratios in brightness temperature
($\propto$ flux $\times$ $\lambda^{2}$).   
The sample contained mostly ultraluminous infrared galaxies (ULIRGs; 
L$_{\rm IR}$ $>$ 10$^{12}$L$_{\odot}$; Sanders et al. 1988a) showing 
luminous buried AGN signatures at other wavelengths, and we found that
their HCN(1--0)/HCO$^{+}$(1--0) brightness-temperature ratios were 
generally higher than starburst galaxies with L$_{\rm IR}$ $<$ 
10$^{12}$L$_{\odot}$.
This result supports the presence of luminous buried AGNs in these
observed ULIRGs.   
However, scenarios that try to explain large HCN(1--0)/HCO$^{+}$(1--0)
brightness-temperatures ratios, without invoking luminous AGNs, have
been present.
Examples include the high turbulence \citep{pap07} or high density
\citep{ima07b} of molecular gas.
In fact, ULIRGs have very dense, highly disturbed molecular gas 
compared to normal LIRGs with L$_{\rm IR}$ $<$ 10$^{12}$L$_{\odot}$ 
(i.e., non-ULIRGs) \citep{sam96,soi00,soi01}.  
Observations of normal LIRGs (non-ULIRGs), with or without buried AGN
signatures, would help elucidate whether the high
HCN(1--0)/HCO$^{+}$(1--0) brightness-temperature ratios in observed
ULIRGs are indeed caused by luminous buried AGNs, or are simply due to
different molecular gas properties in ULIRGs.

In this paper, we present the results of interferometric, simultaneous 
HCN(1--0) and HCO$^{+}$(1--0) observations of such normal LIRGs
(non-ULIRGs), to investigate the HCN(1--0)/HCO$^{+}$(1--0)
brightness-temperature ratios in the nuclear regions.  
This is the final paper of our series of interferometric HCN(1--0) and
HCO$^{+}$(1--0) surveys of LIRGs using NMA \citep{ima04,ink06,in06,ima07b}.
Ancillary ground-based infrared $K$- (1.9--2.5 $\mu$m) and $L$- (2.8--4.1
$\mu$m) band spectra are also utilized to better understand the hidden
energy sources of these millimeter-observed LIRG nuclei. 
Throughout this paper, we adopt H$_{0}$ $=$ 75 km s$^{-1}$ Mpc$^{-1}$, 
$\Omega_{\rm M}$ = 0.3, and $\Omega_{\rm \Lambda}$ = 0.7, to be
consistent with our previously published papers.  

\section{Targets}

We observed the four LIRGs, NGC 2623, Mrk 266, Arp 193, and NGC 1377,
primarily because they were non-ULIRGs with or without buried AGN 
signatures and so were appropriate sources to address the
above-mentioned issue ($\S$1), and because we estimated that we could detect
HCN(1--0) and HCO$^{+}$(1--0) emission with NMA in a reasonable amount
of telescope time. 
Table 1 summarizes the infrared emission properties of these LIRGs. 
An angular scale of 1$''$ corresponds to a physical size of 0.12--0.53
kpc at the redshifts of these LIRGs ($z$ = 0.006--0.028). 

NGC 2623 (z = 0.018) is a LIRG (L$_{\rm IR}$ = 10$^{11.5}$L$_{\odot}$) 
with two prominent merging tails along the north-eastern and south-western
directions \citep{rot04}.
It is optically unclassified \citep{kee84,vei95}; no
obvious AGN signature exists in the optical spectrum. 
Its infrared 5--35 $\mu$m spectrum obtained with {\it Spitzer} IRS is
dominated by polycyclic aromatic hydrocarbon (PAH) emission, as is
usually observed in starburst galaxies \citep{bra06}.
However, high spatial resolution ground-based infrared 10-$\mu$m imaging
observations reveal that the bulk of the 10-$\mu$m emission in NGC 2623
comes from compact nuclear cores with $<$400 pc, the surface
brightness of which can be as high as $\sim$10$^{13}$L$_{\odot}$
\citep{soi01}. This value is close to the maximum found in starburst
phenomena \citep{soi00}.
The 2--10 keV X-ray emission from NGC 2623 is characterized by a hard
spectrum, suggesting that the scattered component of X-ray emission
from a Compton thick (N$_{\rm H}$ $>$ 10$^{24}$ cm$^{-2}$) AGN
dominates the observed 2--10 keV X-ray flux \citep{mai03}. 
The intrinsic AGN luminosity is highly dependent on the unknown value of
the scattering efficiency, and can be high if the efficiency is low. 
The optical non-detection of Seyfert signatures in NGC 2623 suggests
that the putative AGN is a buried one, whose scattering efficiency is
expected to be low \citep{fab02}.
\citet{eva08} performed multi-wavelength observations of NGC 2623, and 
detected the high-excitation forbidden-emission line [NeV]14.3$\mu$m in
the infrared, indicative of narrow line region clouds photoionized
by AGN radiation.  
The intrinsic AGN luminosity can be high because the narrow line
regions are expected to be under-developed in a buried AGN
\citep{idm06,ima07a}.  
The high CO(1--0) flux at the core of NGC 2623 \citep{bry99} suggests
that detection of nuclear HCN(1--0) emission is feasible.

Mrk 266 (NGC 5256; z = 0.028) is a LIRG (L$_{\rm IR}$ =
10$^{11.5}$L$_{\odot}$), consisting of two main merging nuclei 
separated by $\sim$10 arcsec, a south-western (SW) and north-eastern
(NE) nucleus \citep{maz93}.  
Optical spectroscopy by various groups suggested that the SW nucleus
shows clear Seyfert 2 signatures but the NE one does not
\citep{ost83,kol84,maz93,wu98,gon99,ish00}.
However, the most systematic optical spectral classification of galaxies
classified the SW and NE nuclei of Mrk 266 as LINER and Seyfert 2,
respectively \citep{vei95}. 
High spatial resolution X-ray data obtained with {\it Chandra} found 
that the X-ray spectrum of the NE nucleus is characterized by a
heavily-obscured AGN, while that of the SW nucleus consists of an
obscured AGN and a strong starburst component \citep{bra07}.
The detection of a strong 3.3-$\mu$m PAH emission feature \citep{ima02},
a good starburst indicator \citep{moo86,imd00}, supports the presence of
a strong starburst in the SW nucleus. 
The radio 8.44-GHz (3.55 cm) to infrared (1--1000 $\mu$m) luminosity
ratio of Mrk 266 is a factor of 2--3 higher than other LIRGs, suggesting
the presence of a radio-intermediate (or radio-loud) AGN \citep{con91}. 
The radio emission of the NE nucleus is dominated by a spatially compact
($<$0$\farcs$3) component and is much brighter than that of the SW
nucleus \citep{con91}. 
The infrared 5--35 $\mu$m spectrum, taken with {\it Spitzer} IRS, is
dominated by large equivalent-width PAH emission, which is typical of
starburst galaxies \citep{bra06}. 
However, we note that the SW and NE nuclei are not sufficiently covered
with SL (5.2--14.5 $\mu$m) and LL (14--38 $\mu$m) spectra, respectively
\citep{bra06}, possibly missing AGN signatures, given the lack of full
{\it Spitzer} spectral coverage for both nuclei.  
   
Arp 193 (z=0.023) is a LIRG (L$_{\rm IR}$ = 10$^{11.6}$L$_{\odot}$)
with a long narrow tail along the south-eastern direction \citep{rot04}. 
It is classified optically as a LINER \citep{vei95}, showing no
obvious optical AGN signatures. 
The mid-infrared 8--20 $\mu$m dust emission is dominated by a 
spatially-extended component with no prominent compact core
\citep{soi01}. 
The emission surface brightness is $\sim$2 $\times$ 10$^{12}$L$_{\odot}$
\citep{soi01}, which is within the range of starburst phenomena. 
The HCN(1--0) and HCO$^{+}$(1--0) fluxes of Arp 193, measured with
single-dish radio telescopes, are high \citep{sol92,gra08}. 
Millimeter interferometric observations revealed that the CO(1--0) 
emission shows a strong nuclear compact component \citep{dow98,bry99}.
Hence, we expect that detection of nuclear HCN(1--0) emission is
feasible if the nuclear to extended flux ratio for the HCN(1--0) line 
is similar to or higher than that of the CO(1--0) line. 
\citet{pap07} found that the HCN(4--3) to HCN(1--0) flux ratio of Arp
193 is much lower than that of other LIRGs.
The weak high-excitation HCN(4--3) line may suggest that the dominant
energy source is a spatially-extended diffuse one, rather than a
compact high-emission surface-brightness one. 
So far, no obvious AGN signatures have been seen in the Arp 193
observational data \citep{veg08}. 

NGC 1377 (z=0.006) has an infrared luminosity of L$_{\rm IR}$ = 
10$^{10.1}$L$_{\odot}$, and so is not a LIRG in a strict sense. 
This galaxy shows a deficit of 21-cm emission relative to
infrared emission \citep{rou03}. 
Although it is optically unclassified \citep{vei95}, and thus no AGN 
signatures are present in the optical spectrum, the infrared 3--25
$\mu$m spectrum is dominated by a PAH-free continuum 
\citep{ima06,rou06}, typical of AGN-dominated galaxies. 
The 9.7-$\mu$m and 18-$\mu$m silicate dust absorption features are 
extremely strong \citep{rou06}, suggesting that the putative AGN is
deeply buried in dust.  
NGC 1377 is included in our sample to investigate whether even a buried
AGN candidate with low absolute infrared luminosity can have a high
HCN(1--0)/HCO$^{+}$(1--0) brightness-temperature ratio. 
This inclusion is important to test whether the general trend of high
HCN(1--0)/HCO$^{+}$(1--0) brightness-temperature ratios found in ULIRGs
with buried AGN signatures \citep{ink06,ima07b} is indeed due to
AGN-related phenomena, or simply the result of extreme molecular gas
properties in ULIRGs (see $\S$1).  
The CO(1--0) emission is detected with single-dish radio telescope
observations \citep{rou03}.
If a significant fraction of the CO(1--0) emission comes from the
nuclear regions, and unless the nuclear HCN(1--0)/CO(1--0)
brightness-temperature ratio is extremely low, then we estimate that
detection of nuclear HCN(1--0) emission is feasible. 

\section{Observations and Data Reduction}

\subsection{Millimeter Interferometry}

We performed interferometric observations of LIRGs, at 
HCN(1--0) ($\lambda_{\rm rest}$ = 3.3848 mm and $\nu_{\rm rest}$ =
88.632 GHz) and HCO$^{+}$(1--0) ($\lambda_{\rm rest}$ = 3.3637 mm or
$\nu_{\rm rest}$ = 89.188 GHz) lines, using the Nobeyama Millimeter
Array (NMA) at the 
Nobeyama Radio Observatory (NRO) between 2006 December and 2008 March. 
For Mrk 266 and NGC 1377, since no interferometric CO(1--0) maps had
been published, we conducted NMA observations at CO(1--0) 
($\lambda_{\rm rest}$ = 2.6026 mm or $\nu_{\rm rest}$ = 115.271 GHz). 
Table 2 summarizes the detailed observing log. 
The NMA consists of six 10-m antennas and observations were undertaken
using the AB (the longest baseline was 351 m), C (163 m), and D (82 m)
configurations.

The backend was the Ultra-Wide-Band Correlator (UWBC) \citep{oku00}
which can cover 1024 MHz with 128 channels at 8-MHz resolution. 
For HCN(1--0) and HCO$^{+}$(1--0) observations, the central frequency
for each source (Table 2) was set to cover both the redshifted HCN(1--0)
and HCO$^{+}$(1--0) lines simultaneously.  
A bandwidth of 1024 MHz corresponds to $\sim$3500 km s$^{-1}$ for the
redshifted HCN(1--0) and HCO$^{+}$(1--0) lines at $\nu$ $\sim$ 86--89
GHz and $\sim$2700 km s$^{-1}$ for the redshifted CO(1--0) lines 
at $\nu$ $\sim$ 112--115 GHz. 
The fields of view at 86--89 GHz and 112--115 HGz are $\sim$77$''$ and 
$\sim$62$''$ at full-width at half-maximum (FWHM), respectively.
The Hanning window function was applied to reduce side lobes in
the spectra, so that the effective resolution was widened to 16 MHz or
54 km s$^{-1}$ (42 km s$^{-1}$) at $\nu$ $\sim$ 86--89 GHz (112--115 GHz). 

The UVPROC-II package developed at NRO \citep{tsu97} and the AIPS
package of the National Radio Astronomy Observatory were used
for standard data reduction. Corrections for the antenna baselines,
band-pass properties, and time variation in the visibility amplitude
and phase were applied to all of the data (Table 2). 
After discarding a fraction of the data with large phase scatter due to
poor millimeter seeing and clipping a small fraction of data
with unusually high amplitude, the data were Fourier-transformed using a
natural {\it UV} weighting. The flux calibration was made using 
observations of Uranus or appropriate quasars whose flux levels had been
measured at least every month in the NMA observing seasons (Table 2). 
A conventional CLEAN method was applied to deconvolve the synthesized
beam pattern. 
Table 3 summarizes the total net on-source integration times and
synthesized beam patterns. 

\subsection{Ground-based Infrared $K$- (1.9--2.5 $\mu$m) and $L$- 
(2.8--4.1 $\mu$m) band Spectroscopy} 

To put stronger constraints on the hidden energy sources at these
millimeter-observed LIRG nuclei, we performed ground-based infrared 
$K$- (1.9--2.5 $\mu$m) and $L$- (2.8--4.1 $\mu$m) band spectroscopy of NGC
2623, Mrk 266, and Arp 193, as their $L$-band spectra were not
available in the literature.  
NGC 1377 was not observed, because its $L$-band spectrum was given by
\citet{ima06}.  

In short, through infrared $L$-band spectroscopy, we can investigate the
energy sources based on the equivalent width of the 3.3 $\mu$m PAH
emission and the optical depths of absorption features at 
$\lambda_{\rm rest}$ $\sim$ 3.05 $\mu$m by ice-covered dust grains and
at $\lambda_{\rm rest}$ $\sim$ 3.4 $\mu$m by bare carbonaceous dust
grains \citep{imd00,imm03,idm06}. 
A normal starburst galaxy should always show large
equivalent-width 3.3 $\mu$m PAH emission, while a pure AGN produces a
PAH-free continuum \citep{imd00,idm06}.
For absorption features, optical depths have upper limits in a
normal starburst in which stellar energy sources and dust are spatially
well mixed, while the depths can be arbitrarily large in a buried AGN
with a more centrally concentrated energy source geometry than the
surrounding dust \citep{imm03,idm06}. 
In the infrared $K$-band spectra, emission from stars and AGNs is
distinguishable using CO absorption features at $\lambda_{\rm rest}$ =
2.3--2.4 $\mu$m because the CO absorption features are produced by
stars older than 10$^{6}$ yr, but not by AGN-heated hot dust emission.

Infrared $K$- (1.9--2.5 $\mu$m) and $L$- (2.8--4.1 $\mu$m) band spectra
were taken using SpeX \citep{ray03} attached to the IRTF 3-m telescope
atop Mauna Kea, Hawaii, on 2008 Apr 19 and 20 (UT). The 1.9--4.2 $\mu$m
cross-dispersed mode was employed, so that $K$- and $L$-band spectra
were obtained simultaneously. 
We chose a narrow 0\farcs8-wide slit to pinpoint the infrared continuum
emission peaks of individual LIRG nuclei where the putative AGN may be
located.  
The resulting spectral resolution using this slit is R $\sim$ 1000 in
the $K$- and $L$-bands.

Photometric conditions persisted throughout the observations. The
seeing at $K$ was measured in the range 0$\farcs$6--0$\farcs$8 in
FWHM. A standard telescope nodding technique (ABBA pattern) with a throw
of 7$\farcs$5 was employed along the slit. 
Each exposure was 15 sec, and two coadds were made at each position. 
The telescope tracking was monitored with the SpeX infrared slit-viewer.
Table 4 summarizes the detailed observing log. 

Appropriate standard stars (Table 4), with an airmass
difference of $<$0.1 for individual LIRGs, were observed to correct for the
wavelength-dependent transmission of Earth's atmosphere.  
The $K$- and $L$-band magnitudes of the standard stars were estimated
based on their $V$-band (0.6 $\mu$m) magnitudes, and $V-K$ and $V-L$ 
colors, of the corresponding stellar types \citep{tok00}. 
The estimated $K$-band magnitudes agree very well with the 2MASS
measurements \citep{skr06}.

Standard data reduction procedures were employed using IRAF
\footnote{
IRAF is distributed by the National Optical Astronomy Observatories,
operated by the Association of Universities for Research
in Astronomy, Inc. (AURA), under cooperative agreement with the
National Science Foundation.}. 
Initially, frames taken with an A (or B) beam were subtracted from
frames subsequently taken with a B (or A) beam, and the resulting
subtracted frames were added and divided by a spectroscopic flat
image.  Then, bad pixels and pixels hit by cosmic rays were replaced
with interpolated values from surrounding pixels.  Finally the
spectra of LIRG nuclei and standard stars were extracted by
integrating signals over 1$\farcs$8--2$\farcs$7, depending on actual
signal profiles. 
Wavelength calibration was performed based on standard star spectra 
which reflect the wavelength-dependent nature of Earth's atmospheric
transmission curve with high signal-to-noise ratios. 
The LIRG nuclei spectra were divided by the observed spectra of
standard stars, and then multiplied by the spectra of blackbodies
with temperatures appropriate to individual standard stars (Table 4). 

Flux calibration was done based on signals of LIRGs and
standard stars detected inside our slit spectra. 
Since we employed a narrow 0$\farcs$8-wide slit, our spectra are
sensitive only to spatially-compact emission, and miss a 
spatially-extended component.  
To reduce data point scatter, appropriate binning of spectral
elements was performed.

\section{Results}

\subsection{Millimeter Interferometric Data}

For NGC 2623, Mrk 266, and Arp 193, our spectra at the HCN(1--0) or 
HCO$^{+}$(1--0) emission peaks show that the flux levels between these
lines are above zero, indicating that continuum emission is 
present.  
We combined data points that are unaffected by these lines and made
interferometric maps of the continuum emission. 
Figure 1 presents the contours of the continuum emission for NGC 2623, 
Mrk 266, and Arp 193, in the vicinity of the redshifted HCN(1--0) and
HCO$^{+}$(1--0) emission at 86--89 GHz. 
For Arp 193, the continuum emission ($\sim$15 mJy) is clearly
($>$8$\sigma$) detected. 
For NGC 2623, and Mrk 266 SW and NE, signs of continuum emission are
marginally seen ($\sim$3$\sigma$). 
Continuum emission was not detected in the HCN(1--0)/HCO$^{+}$(1--0)
data of NGC 1377, or the CO(1--0) data of Mrk 266 and NGC 1377.  

Figure 2 displays integrated intensity maps of the HCN(1--0) and
HCO$^{+}$(1--0) emission of the observed four LIRGs. 
Figure 3 presents the interferometric maps of CO(1--0) emission for Mrk 
266 and NGC 1377. 
Since LIRGs contain highly concentrated nuclear molecular gas
\citep{sam96}, the fraction of high-density molecular gas (n$_{\rm H}$
$>$ 10$^{4}$ cm$^{-3}$) is expected to increase \citep{gao04}.  
Such molecular gas is better probed with HCN(1--0) and HCO$^{+}$(1--0)
rather than the widely used CO(1--0) because the dipole moments of 
HCN and HCO$^{+}$ ($\mu$ $>$ 3 debye) are much larger than CO 
($\mu$ $\sim$ 0.1 debye; Botschwina et al. 1993; Millar et al. 1997).
Hence, the maps in Fig. 2 reflect the spatial distribution of
the dense molecular gas phase.

Mrk 266 has a double-nucleus morphology. 
The CO(1--0) emission is sufficiently bright to investigate its velocity
information for each individual nucleus, as shown in Figure 4.
In this CO(1--0) channel map (Fig. 4), the NE nucleus shows a 
CO(1--0) emission peak at lower frequency (112.08--112.11 GHz) than the
SW nucleus (112.17--112.19 GHz), suggesting that the NE nucleus is more
redshifted than the SW nucleus.  
Previously obtained optical spectra also show that the NE nucleus is more
redshifted than the SW nucleus (20--130 km s$^{-1}$)
\citep{maz93,kol84,kim95}.  
  
Figure 5 shows spectra around the HCN(1--0) and HCO$^{+}$(1--0)
lines at the nuclear positions of the observed four LIRGs.
Spectra around the CO(1--0) line for Mrk 266 SW and NE, and NGC
1377 are presented in Figure 6. 
For NGC 1377, HCN(1--0) emission might be marginally detected in the
integrated intensity map close to the CO(1--0) 
emission peak (Figs. 2 and 3). 
In the spectrum at this peak position, five successive spectral elements
at the expected HCN(1--0) frequency show flux excess, compared to the
surrounding pixels (Fig. 5, lower-left). 
For reference, a spectrum with less binning, despite being slightly noisier,
also displays signatures of HCN(1--0) emission (Fig. 5, lower-right).    
We thus regard the HCN(1--0) emission signature as real in NGC 1377.

Figure 7 presents Gaussian fits to the detected HCN(1--0),
HCO$^{+}$(1--0), and CO(1--0) lines.
In the HCN(1--0)/HCO$^{+}$(1--0) spectra, the central velocity and line
width of the Gaussian components are determined independently for both lines.
For NGC 1377, although HCN(1--0) and CO(1--0) emission are seen in
similar positions (Figs. 2 and 3), their velocity profiles are 
significantly different (Fig. 7).
It may be that the HCN(1--0) line reflects high density molecular gas in the
central nuclear region of NGC 1377, while CO(1--0) emission probes the 
surrounding diffuse molecular gas toward the nuclear direction. 
We should take care interpreting the HCN(1--0)/CO(1--0)
brightness-temperature ratio of NGC 1377 because HCN(1--0) and CO(1--0)
lines may probe physically unrelated molecular gas components. 

Table 5 summarizes the derived Gaussian fitting parameters.
The derived fluxes of HCN(1--0) and HCO$^{+}$(1--0), and their 
brightness-temperature ratios at the nuclear peak positions (i.e.,
spatially unresolved component) are also summarized in Table 5.
For NGC 2623 and Arp 193, where continuum emission is subtracted, 
the integrated HCN(1--0) and HCO$^{+}$(1--0) fluxes are also estimated 
from the peak contours of the integrated intensity maps in Fig. 2.  
The values estimated in this way agree with those based on the Gaussian
fits to within 30\%.  
Table 5 also includes CO(1--0) fluxes at the nuclear peaks, derived from
our NMA data (Mrk 266 and NGC 1377) or taken from the literature (NGC
2623 and Arp 193). 

The CO(1--0) emission at Mrk 266 SW clearly displays
spatially-extended structures (Fig. 3). 
The total CO(1--0) flux, including this spatially-extended component, 
is estimated to be 162 Jy km s$^{-1}$.

CO(1--0) fluxes of Mrk 266 and NGC 1377, and HCN(1--0) and
HCO$^{+}$(1--0) fluxes of Arp 193, have been measured using single-dish
radio telescopes \citep{san86,rou03,gra08}.
Table 6 compares these single-dish measurements with our NMA interferometric
data. 
For Mrk 266 CO(1--0) emission, our interferometric flux measurements 
provide good agreement, within $\sim$15\%, to old single-dish measurements. 
For Arp 193, our HCN(1--0) flux is more comparable to the latest
IRAM 30m measurement \citep{gra08}, but is a factor of two smaller than
the old IRAM one \citep{sol92}, whose high flux measurement was noted by
\citet{gra08}.  
For Arp193 HCO$^{+}$(1--0) emission, our measurement agrees with the 
latest IRAM 30m data \citep{gra08}. 
For these lines, the bulk of the emission should be covered in our
interferometric data.
For the nearest source, NGC 1377, however, our CO(1--0) flux measurement
is more than a factor of four smaller than the SEST telescope 
measurement, suggesting that our interferometric data miss a spatially
very extended component, and detect only nuclear compact molecular gas.  

\subsection{Ground-based Infrared Spectra}

Figure 8 presents ground-based infrared $K$- (1.9--2.5 $\mu$m) and $L$-
(2.8--4.1 $\mu$m) band spectra of NGC 2623, Mrk 266 SW and NW, and Arp 193.
$K$-band spectra with narrower wavelength coverage are available 
for NGC 2623, Arp 193, and Mrk 266 SW \citep{rid94,gol97}. 
For Mrk 266 SW, an $L$-band spectrum is also available \citep{ima02}. 
A $K$-band spectrum of Mrk 266 NE, and $L$-band spectra of 
NGC 2623, Mrk 266 NE, and Arp 193 are first presented in this paper.
For Mrk 266, this is the first infrared spectrum that clearly resolves 
emission from the Mrk 266 NE and SW nuclei, allowing 
separate investigation of the energy sources of individual nuclei for
the first time.  

\subsubsection{Continuum Flux Level}

For NGC 2623, and Mrk 266 NE and SW, the $K$-band flux levels of
our SpeX spectra (0$\farcs$8 slit width) are a factor of 1.7--2.7 smaller than
the $K$-band photometric measurements with a 5'' aperture \citep{car90}.
For NGC 2623, Mrk 266 SW, and Arp 193, the $K$-band flux levels of
our SpeX spectra (0$\farcs$8 slit width) are 30--50\% smaller than those
measured in 3'' $\times$ 9'' aperture spectra \citep{gol97}.
The difference is greatest for Mrk 266 SW, whose spatially-extended 
$K$-band emission is strongest in our spectra along the slit direction.
The smaller flux levels in our spectra are reasonable, given that our
smaller aperture misses a spatially extended emission component.

\subsubsection{Hydrogen Recombination Lines}

Strong Pa$\alpha$ (1.87 $\mu$m) and Br$\alpha$ (4.05 $\mu$m) emission
lines are visible in some spectra. Enlarged spectra around these strong
emission lines are shown in Figure 9. 
In the $K$-band spectra, Br$\gamma$ (2.17 $\mu$m) and H$_{2}$(1--0) S(1)
(2.12 $\mu$m) emission lines are detectable. 
Table 7 summarizes the strengths and line widths of these emission
lines.  
If emission from the broad line regions close to the
central SMBH in an AGN is strong and only modestly obscured, then 
we may be able to detect broad (FWHM $>$ 1500 km s$^{-1}$) emission line
components in the infrared, where dust extinction is much smaller than
the optical, as was seen in optical Seyfert 2 galaxies
\citep{hil96,vei97,lut02}. 
However, we see no obvious signatures of broad line components. 
Since buried AGNs tend to contain a larger amount of obscuring dust than
Seyfert 2 AGNs \citep{idm06,ima07a,ima08}, detection of the broad emission
line components would be more difficult, even if the buried AGNs
emit intrinsically luminous broad emission lines. 

In Table 7, since our small aperture spectra miss spatially-extended
emission, the absolute flux is not meaningful. 
Only the flux ratios between different emission lines measured with the
same apertures will be used in our discussion.

\subsubsection{PAH Emission}

The 3.3-$\mu$m PAH emission is clearly seen in the $L$-band spectra of
all the observed LIRGs. 
To estimate the 3.3-$\mu$m PAH emission strength, we make
the reasonable assumption that the profiles of the 3.3-$\mu$m PAH
emission in these LIRGs are similar to those of Galactic star-forming
regions and nearby starburst galaxies (type-1 sources in Tokunaga et
al. 1991), following \citet{idm06}. The adopted profile reproduces the
observed 3.3-$\mu$m PAH emission features of the LIRGs reasonably well. 
Table 8 summarizes the fluxes, luminosities, and rest-frame equivalent
widths of the 3.3 $\mu$m PAH emission feature.
Only the equivalent width is not significantly affected by flux loss in
our small-aperture spectra, and so will be used in our discussion.

\subsubsection{CO Absorption}

Figure 10 shows enlarged spectra in the longer wavelength portion of the
$K$-band spectra.
All spectra show spectral gaps in the continuum at
$\lambda_{\rm obs} >$ 2.35 $\mu$m in the observed frame.
We attribute the gaps to CO absorption features at $\lambda_{\rm rest}$
= 2.31--2.4 $\mu$m produced by stars older than 10$^{6}$ yr 
\citep{oli99,iva00,ia04,imw04}. 
To estimate the CO absorption strengths, we adopt the spectroscopic CO
index (CO$_{\rm spec}$) defined by \citet{doy94} and follow the
procedures previously applied to other LIRGs by ourselves 
\citep{ima04,in06}. 
Power-law continuum levels (F$_{\rm \lambda}$ = $\alpha \times
\lambda^{\beta}$), shown as solid lines in Fig. 10, are determined
using data points at $\lambda_{\rm rest}$ = 2.05--2.29 $\mu$m, excluding
obvious emission lines. 
The derived CO$_{\rm spec}$ values are summarized in column 5 of Table 8.

\citet{rid94} also estimated CO$_{\rm spec}$ = 0.24$\pm$0.03 for NGC 2623,
with a 2$\farcs$7 aperture, which agrees with our measurement of 
CO$_{\rm spec}$ = 0.27 (Table 8) to within uncertainty.
\citet{gol97} measured the photometric CO index inside 3'' $\times$ 9''
apertures and estimated CO$_{\rm ph}$ = 0.23, 0.16, and 0.18 for NGC
2623, Mrk 266SW, and Arp 193, respectively. 
Using the formula CO$_{\rm spec}$ = 1.46 $\times$ CO$_{\rm ph}$ $-$ 
0.02 \citep{gol97}, these CO$_{\rm phot}$ values are converted to 
CO$_{\rm spec}$ = 0.32, 0.21, and 0.24 for NGC 2623, Mrk 266SW, and Arp
193, respectively, which are slightly (0.02--0.05) larger than 
our measurements (Table 8).

\section{Discussion}

\subsection{Comparison of HCN(1--0)/HCO$^{+}$(1--0)
Brightness-Temperature Ratios with Other Galaxies}

Figure 11 plots the HCN(1--0)/HCO$^{+}$(1--0) and HCN(1--0)/CO(1--0)
brightness-temperature ratios for the four LIRGs. 
Previously obtained data points of nearby LIRGs
\citep{ima04,ink06,in06,ima07b}, starbursts, and Seyfert galaxies
\citep{koh05} are also plotted. 
As stated by \citet{ink06}, the HCN(1--0)/HCO$^{+}$(1--0)
brightness-temperature ratios in the ordinate are mainly used in our
discussions for the following reasons.  
First, since HCN(1--0) and HCO$^{+}$(1--0) have similarly high dipole
moments ($\mu$ $>$ 3 debye; Botschwina et al. 1993; Millar et al. 1997),
it is very likely that similar high-density molecular gas is probed
with both lines.  
Next, as both the HCN(1--0) and HCO$^{+}$(1--0) lines are observed
simultaneously with the same NMA array configuration, their beam
patterns are virtually identical.
Finally, both HCN(1--0) and HCO$^{+}$(1--0) fluxes are measured
at the same time with the same receiver and same correlator unit, under
the same weather conditions, so that possible {\it absolute} flux
calibration uncertainties of interferometric data do not propagate to
the ratio, which is dominated by statistical noise
and fitting errors (see Fig. 7).  

In Fig. 11, we note the following points: 
first, NGC 1377 exhibits a high HCN(1--0)/HCO$^{+}$(1--0)
brightness-temperature ratio, as seen in AGN-dominated galaxy nuclei. 
Second, the Mrk 266 NE nucleus shows a significantly higher ratio than the SW
nucleus. 
Finally, the HCN(1--0)/HCO$^{+}$(1--0) brightness-temperature ratios of
Mrk 266 SW and Arp 193 are low, as observed in starburst galaxies, and 
that of NGC 2623 is higher than Mrk 266 SW and Arp 193. 

\subsection{Buried AGN Signatures in Individual Objects} 

In this subsection, we look for buried AGN signatures in the infrared
spectra of individual LIRG nuclei and then investigate their 
HCN(1--0)/HCO$^{+}$(1--0) brightness-temperature ratios. 
In doing so, we have to account for the fact that starburst
activity surrounds the central compact buried AGN, so starburst
emission is less obscured by gas and dust than the buried AGN emission.
The contribution from the buried AGN to the observed infrared flux can
be small, even if the buried AGN is intrinsically luminous and is
energetically important, due to dust extinction.  
Thus, we need to examine the infrared spectra carefully. 

\subsubsection{Mrk 266 NE}

Although optical spectroscopy by various groups has failed to find AGN 
signatures in the Mrk 266 NE nucleus
\citep{ost83,kol84,maz93,wu98,gon99,ish00}, buried AGN signatures are
discernible in our infrared spectra. 
First, Mrk 266 NE shows a high H$_{2}$(1--0) S(1) to Br$\gamma$ flux ratio, 
$>$ 1 (Table 7). Observationally, starburst galaxies show 
H$_{2}$(1--0) S(1) to Br$\gamma$ flux ratios of $<$1 \citep{mou92}, while 
high H$_{2}$(1--0) S(1) to Br$\gamma$ flux ratios ($>$1) are found in optical
AGNs (= Seyfert galaxies) \citep{mou92}, as well as buried AGN candidates
\citep{ima04}.  
Next, the rest-frame equivalent width of the 3.3-$\mu$m PAH emission 
of Mrk 266 NE (EW$_{\rm 3.3PAH}$ $\sim$ 30 nm; Table 8) is significantly
smaller than that observed in starburst galaxies ($\sim$100 nm; Moorwood
1986; Imanishi \& Dudley 2000), suggesting that a PAH-free AGN-heated
hot dust continuum contributes significantly to the observed 3--4 $\mu$m
flux of Mrk 266 NE. 

The HCN(1--0)/HCO$^{+}$(1--0) brightness-temperature ratio of Mrk 266 NE
is also higher than that of the Mrk 266 SW nucleus. 
A putative buried AGN, suggested from X-ray data \citep{bra07}, may
enhance the HCN(1--0) emission in Mrk 266 NE.

\subsubsection{NGC 1377}

The HCN(1--0)/HCO$^{+}$(1--0) brightness-temperature ratio of NGC 1377
is high, as found in AGNs.
The infrared $L$-band spectrum is also dominated by PAH-free continuum
emission \citep{ima06}, as usually seen in AGNs.
The strong 9.7 $\mu$m silicate dust absorption feature detected in the 
{\it Spitzer} IRS infrared 5--35 $\mu$m spectrum \citep{rou06} is also
incompatible with a normal starburst, where stellar energy sources and
dust are spatially well mixed.  It requires a buried AGN-type
centrally concentrated energy source geometry \citep{ima07a}.  
These overall observational results are naturally explained by 
the presence of a luminous buried AGN in NGC 1377.

A PAH-free continuum and strong 9.7-$\mu$m silicate dust absorption
could be explained by an exceptionally centrally concentrated extreme 
starburst whose emitting volume is predominantly occupied with
HII-regions, with virtually no molecular gas and photo-dissociation
regions (Fig. 1e of Imanishi et al. 2007a).
Unlike ULIRGs with L$_{\rm IR}$ $>$ 10$^{12}$L$_{\odot}$, the absolute
infrared luminosity of NGC 1377 is only $\sim$10$^{10.1}$L$_{\odot}$, 
which could be accounted for by a small number of super star cluster
whose emitting size is very small ($<<$100 pc) and emission surface
brightness is high \citep{gor01}. 
In fact, \citet{rou03} preferred such a compact starburst (super star
cluster) scenario, but ruled out the possibility of an energetically
important buried AGN, based on the following arguments:  
(1) estimated SMBH mass is too small to account
for the luminosity of NGC 1377 with AGN activity, and 
(2) high excitation forbidden emission lines, usually seen in Seyfert
galaxies, are not detected. 

Regarding the first point, the SMBH mass was estimated to be $<$2
$\times$ 10$^{5}$M$_{\odot}$, from the observed radio 20-cm (1.5-GHz)
continuum flux and the small line width of the infrared H$_{2}$(1--0)
S(1) emission line (FWHM $<$ 25 km s$^{-1}$) \citep{rou03}. 
The measured line width of millimeter CO(1--0) lines in our NMA spectrum
(Figure 6) is $\sim$100 km s$^{-1}$ in FWHM. 
After correction for the velocity resolution of NMA data ($\sim$42 km
s$^{-1}$; $\S$3.1), we obtain the intrinsic CO(1--0) line width of 
$\sim$90 km s$^{-1}$ in FWHM, which is similar to the SEST 15-m
telescope measurement \citep{rou03}.  
Assuming the CO(1--0) line width and SMBH mass relation given by
\citet{shi06},  
the measured CO(1--0) line width provides a similar SMBH mass of
$\sim$2 $\times$ 10$^{5}$M$_{\odot}$. 
The Eddington luminosity of this SMBH mass is $\sim$7 $\times$ 
10$^{9}$L$_{\odot}$. 
This is only slightly below the observed infrared luminosity
($\sim$1 $\times$ 10$^{10}$L$_{\odot}$).
In a buried AGN, the infrared luminosity should be
comparable to the bolometric luminosity, because almost all of the
energetic radiation from the central AGN is absorbed by the
surrounding dust and re-emitted as infrared thermal dust emission. 
Thus, even if current constraints on the supermassive black hole
mass are accurate, it is still possible that the bulk of the observed 
luminosity in NGC 1377 comes from buried AGN activity. 

However, uncertainties in the estimated SMBH mass in NGC 1377 could exist.
First, the observed 20-cm flux from a buried AGN can be severely
attenuated by free-free absorption, possibly leading to an underestimate
of the SMBH mass. 
Second, the line width of H$_{2}$(1--0) S(1) emission is not well
calibrated for an SMBH mass estimate.
Finally, in the small line width range of millimeter CO(1--0) 
($<$100 km s$^{-1}$), the SMBH mass estimated from CO(1--0) line width 
yields systematically smaller results than other measurements
\citep{shi06}. 
Hence, the actual SMBH mass in NGC 1377 could be higher than above
estimates. 

Regarding the second point, the non-detection of high excitation
emission lines is a natural consequence of a buried AGN because the
central AGN is obscured by dust and gas along virtually all directions
at the inner part ($<$1 pc), producing virtually no narrow line regions
(= the main emitting sources of high excitation lines).  

Therefore, none of the currently available observational results
preclude the presence of a luminous buried AGN in NGC 1377. 
Strong H$_{2}$ emission is observed in NGC 1377 \citep{rou06}.
In general, H$_{2}$ emission is stronger in AGNs than starburst galaxies
\citep{mou92}. 
Although a major galaxy merger could produce strong H$_{2}$ emission
\citep{van93}, NGC 1377 shows no sign of a major merger \citep{rou06}.
Energy source obscuration for NGC 1377 is extremely high
\citep{rou06}, and the observed low radio to infrared luminosity ratio
\citep{rou03} could be explained by severe flux attenuation of the radio
20-cm (1.5 GHz) emission by free-free absorption.

\subsubsection{NGC 2623}

The EW$_{\rm 3.3PAH}$ and CO$_{\rm spec}$ values, and the H$_{2}$(1--0)
S(1) to Br$\gamma$ flux ratio of NGC 2623 are all within starburst range.
The HCN(1--0)/HCO$^{+}$(1--0) brightness-temperature ratios are slightly
higher than Arp 193 and Mrk 266 SW, which may be due to an HCN(1--0)
emission enhancement by an X-ray detected buried AGN (see $\S$2). 

\subsubsection{Mrk 266 SW and Arp 193}

The large EW$_{\rm 3.3PAH}$ and CO$_{\rm spec}$ values, the small
H$_{2}$(1--0) S(1) to Br$\gamma$ flux ratios, and the small
HCN(1--0)/HCO$^{+}$(1--0) brightness-temperature ratios of Mrk 266 SW
and Arp 193 are all explained by starburst activity only, with no
significant AGN contribution required. 

\subsection{Interpretation of High HCN(1--0)/HCO$^{+}$(1--0) 
Brightness-Temperature Ratios in Buried AGN Candidates} 

One natural explanation for the high HCN(1--0)/HCO$^{+}$(1--0)
brightness-temperature ratios found in AGNs is an HCN abundance
enhancement. 
If molecular gas consists of small dense
gas clumps with low volume filling factor, as widely supported from 
observations \citep{sol87}, an enhanced HCN abundance can result in
higher HCN(1--0) flux regardless of whether the HCN(1--0) emission is
optically thin or thick \citep{ima07b}. 
Several chemical calculations of HCN abundance in molecular gas
around UV- and X-ray emitting energy sources have been published 
\citep{mei06,lin06}. 
Although an HCN abundance enhancement around an X-ray emitting energy
source (i.e., AGN) is predicted in some parameter ranges that are
realistic for molecular gas around buried AGNs in LIRGs (high-F$_{\rm X}$
range of Table 3 in Meijerink et al. 2006), an HCN abundance 
{\it decrease} is suggested in other reasonable parameter ranges
\citep{mei06}. 
The chemical calculation results are highly dependent on parameters, and
it is currently unclear whether an HCN abundance is indeed enhanced in
molecular gas in the close vicinity of buried AGNs in LIRG nuclei.  

The second possible explanation for the strong HCN(1--0) emission in
AGNs is an infrared radiative pumping scenario \citep{aal95,gar06,wei07}.
It is usually assumed that molecular gas is excited by collision.
However, the HCN molecule has a line at infrared 14 $\mu$m that can be
excited by absorbing photons at $\lambda \sim$ 14 $\mu$m.
HCN(1--0) emission in the millimeter
wavelength range could be enhanced through a cascade process.
Since the emission surface brightness of an AGN is high, 
much of the surrounding dust is heated to 
several 100 K, producing strong mid-infrared 10--20 $\mu$m
emission ($\S$1). 
Hence, this infrared radiative pumping scenario for HCN could
work effectively in AGNs.  

HCO$^{+}$ also has a line at infrared 12 $\mu$m, so that the infrared
radiative pumping mechanism could work in a similar way.  
Observationally, the 14 $\mu$m HCN absorption features are detected in 
highly obscured LIRGs \citep{lah07}, but the 12 $\mu$m HCO$^{+}$  
absorption features are not \citep{far07}. 
It may be that this infrared radiative pumping scenario is working more
effectively for HCN than HCO$^{+}$, possibly enhancing the
HCN(1--0)/HCO$^{+}$(1--0) brightness-temperature ratios in AGNs.

To test this infrared radiative pumping scenario, in Figure 12 we compare the 
observed HCN(1--0)/HCO$^{+}$(1--0) brightness-temperature ratio with
the infrared emission surface brightness estimated by \citet{soi00,soi01}
and \citet{eva03}.
We use the infrared emission surface brightness, rather than infrared
luminosity \citep{gra06}, because the bulk of the observed normal LIRGs
with $<$10$^{12}$L$_{\odot}$ (i.e., non-ULIRGs) are chosen because they
display luminous buried AGN signatures. 
Although \citet{gra08} found an increasing trend of
HCN(1--0)/HCO$^{+}$(1--0) brightness-temperature ratios with increasing
galaxy infrared luminosities, our heterogeneous sample selection of
normal LIRGs (non-ULIRGs) could be biased to AGN candidates (= strong
HCN emitters) and artificially eliminate this trend. 
The infrared radiative pumping scenario can, in principle, enhance the
HCN(1--0)/HCO$^{+}$(1--0) brightness-temperature ratio, even for a low 
absolute infrared luminosity galaxy (e.g. NGC 1377), if infrared
emission surface brightness is high. 

In Fig. 12, we may see a weak correlation but the scatter is large and
the total sample size is small. 
Based on our current dataset, we cannot determine
whether the infrared
radiative pumping scenario is indeed at work in buried AGNs at LIRG's 
nuclei.  
Further detailed theoretical calculations that realistically
incorporate the actual energy level populations of HCN and HCO$^{+}$, as
well as the clumpy structure of molecular gas around an AGN
\citep{yam07}, are needed to properly interpret the origin of strong
HCN(1--0) emission in AGNs.   
Combination with higher transition lines in the submillimeter wavelength
range \citep{wil08} will also help better understand the physical 
properties of molecular gas in LIRG's nuclei.

\section{Summary}

We presented the results of millimeter interferometric simultaneous
HCN(1--0) and HCO$^{+}$(1--0) observations of four LIRGs using NMA.
When combined with previously observed LIRGs, 
ours is the largest LIRG interferometric HCN(1--0) and HCO$^{+}$(1--0) survey. 
From our interferometric data, we extracted the HCN(1--0) and
HCO$^{+}$(1--0) fluxes at the nuclei, where putative AGNs
are expected to be present. 
We derived the HCN(1--0)/HCO$^{+}$(1--0) brightness-temperature ratios 
of the observed LIRG nuclei and compared them with the ratios found in
AGNs and starburst galaxies. 
The main result of this paper is the confirmation of the
previously discovered trend that LIRGs with luminous buried AGN signatures at
other wavelengths tend to show higher HCN(1--0)/HCO$^{+}$(1--0)
brightness-temperature ratios than those without.
This reinforces the utility of the HCN(1--0)/HCO$^{+}$(1--0)
brightness-temperature ratio as a powerful observational tool for  
discovering elusive AGNs deeply buried in dust and molecular gas.

\acknowledgments

We are grateful to the NRO staff for their continuous efforts to make 
our NMA observations possible.
We also thank D. Griep for his support during our IRTF observing runs,
and the anonymous referee for his/her useful comments. 
M.I. is supported by Grants-in-Aid for Scientific Research (19740109). 
Y.T. is financially supported by the Japan Society for the Promotion of
Science (JSPS) for Young Scientists. 
NRO is a branch of the National Astronomical Observatory, National
Institutes of Natural Sciences, Japan.
This study utilized the SIMBAD database, operated at CDS,
Strasbourg, France, and the NASA/IPAC Extragalactic Database
(NED) operated by the Jet Propulsion Laboratory, California
Institute of Technology, under contract with the National Aeronautics
and Space Administration.
This publication makes use of data products from the Two Micron All Sky
Survey, which is a joint project of the University of Massachusetts and
the Infrared Processing and Analysis Center/California Institute of
Technology, funded by the National Aeronautics and Space Administration
and the National Science Foundation.

\clearpage

\begin{deluxetable}{lcccrrcl}
\tabletypesize{\scriptsize}
\tablecaption{Detailed information on the observed LIRGs 
\label{tab1}}
\tablewidth{0pt}
\tablehead{
\colhead{Object} & \colhead{Redshift}   & 
\colhead{$f_{\rm 12}$}  & \colhead{$f_{\rm 25}$}  & 
\colhead{$f_{\rm 60}$}  & \colhead{$f_{\rm 100}$}  & 
\colhead{log $L_{\rm IR}$ (log $L_{\rm IR}$/$L_{\odot}$)} &
\colhead{Far-infrared} \\ 
\colhead{} & \colhead{}   & \colhead{(Jy)} & \colhead{(Jy)}  & \colhead{(Jy)} 
& \colhead{(Jy)}  & \colhead{(ergs s$^{-1}$)} & \colhead{Color} \\
\colhead{(1)} & \colhead{(2)} & \colhead{(3)} & \colhead{(4)} & 
\colhead{(5)} & \colhead{(6)} & \colhead{(7)} & \colhead{(8)}
}
\startdata
NGC 2623 (Arp 243) & 0.018 & 0.21 & 1.74 & 23.13 & 27.88 & 45.1 (11.5) &
0.08 (cool) \\ 
Mrk 266 (NGC 5256) & 0.028 & 0.23 & 0.98 & 7.34 & 11.07 & 45.0 (11.5) &
0.13 (cool) \\  
Arp 193 (IC 883, UGC 8387) & 0.023 & 0.26 & 1.36 & 15.44 & 25.18 & 45.2
(11.6) & 0.09 (cool) \\  
NGC 1377 & 0.006 & 0.44 & 1.81 & 7.25 & 5.74 & 43.7 (10.1) & 
0.25 (warm) \\ 
\enddata

\tablecomments{
Col.(1): Object name.
Col.(2): Redshift.
Cols.(3)--(6): f$_{12}$, f$_{25}$, f$_{60}$, and f$_{100}$ are
{\it IRAS FSC} fluxes at 12, 25, 60, and 100 $\mu$m, respectively.
Col.(7): Decimal logarithm of the infrared (8$-$1000 $\mu$m) luminosity
in ergs s$^{-1}$ calculated as follows:
$L_{\rm IR} = 2.1 \times 10^{39} \times$ D(Mpc)$^{2}$
$\times$ (13.48 $\times$ $f_{12}$ + 5.16 $\times$ $f_{25}$ +
$2.58 \times f_{60} + f_{100}$) ergs s$^{-1}$
\citep{sam96}.
The values in parentheses are the decimal logarithms of the infrared
luminosities in units of solar luminosities.
Col.(8): {\it IRAS} 25 $\mu$m to 60 $\mu$m flux ratio
(f$_{25}$/f$_{60}$). LIRGs with f$_{25}$/f$_{60}$ $<$ ($>$) 0.2 are
classified as cool (warm) \citep{san88b}.
}

\end{deluxetable}

\begin{deluxetable}{lclccll}
\tabletypesize{\scriptsize}
\tablecaption{Observing log of NMA millimeter interferometric
observations \label{tab2}}  
\tablewidth{0pt}
\tablehead{
\colhead{Object} & \colhead{Array} & \colhead{Observing Date} &
\colhead{Central} & \colhead{Phase} & \colhead{Band-pass} & \colhead{Flux} \\ 
\colhead{} & \colhead{configuration} & \colhead{(UT)} &
\colhead{frequency} & \colhead{calibrator} & \colhead{calibrator} &
\colhead{calibrator} \\  
\colhead{(1)} & \colhead{(2)} & \colhead{(3)} & \colhead{(4)} &
\colhead{(5)} & \colhead{(6)} & \colhead{(7)}
}
\startdata
NGC 2623 & C & 2007 Dec 13, 14, 15, 16 & 87.30 & 0851+202 & 3C 84 &
3C 345 \\ \hline
Mrk 266  & C & 2008 Jan 8, 9, 10 & 86.52 & 1418+546 & 3C 84 & 1749+096 \\  
     & D & 2008 Feb 20, 21, 22, Mar 4, 5 & & & & \\ 
& C (CO) & 2007 Dec 19, 23 & 112.43  & 1418+546 & 3C 84 & 1749+096 \\
\hline
Arp 193 & C & 2007 Dec 20 & 86.89 & 1308+326 & 3C 84 & 1749+096 \\ 
                & D  & 2008 Feb 18, 19 & & & & \\ \hline
NGC 1377 & D & 2006 Dec 12, 13  & 88.47 & 0420$-$014 & 3C 454.3 & Uranus \\
 & AB & 2007 Jan 27, 28 & & & & \\
 & C & 2007 Dec 13, 14, 15, 16, 19, 20 & & & &  \\
 & C (CO) & 2007 Dec 10, 11 & 114.95 & 0420$-$014 & 3C 454.3 & Uranus \\
\hline 
\enddata

\tablecomments{
Col.(1): Object name.
Col.(2): NMA array configuration.
            The mark ``(CO)'' means CO(1--0) observations, which were
            executed only for Mrk 266 and NGC 1377.  
Col.(3): Observing date in UT.
Col.(4): Central frequency used for the observations.
Col.(5): Object name used as a phase calibrator.
Col.(6): Object name used as a band-pass calibrator. 
Col.(7): Object name used as a flux calibrator. 
}

\end{deluxetable}

\begin{deluxetable}{llccc}
\tablecaption{Parameters of final NMA maps \label{tab3}}  
\tablewidth{0pt}
\tablehead{
\colhead{Object} & \colhead{Line} & \colhead{On source} & 
\colhead{Beam size} & \colhead{Position angle of} \\  
\colhead{} & \colhead{} & \colhead{integration (hr)} & 
\colhead{(arcsec $\times$ arcsec)} & 
\colhead{the beam ($^{\circ}$)} \\
\colhead{(1)} & \colhead{(2)} & \colhead{(3)} & \colhead{(4)} & \colhead{(5)}
}
\startdata
NGC 2623 & HCN/HCO$^{+}$ & 10 & 5.4 $\times$ 4.2 & $-$48.4 \\
Mrk 266  & HCN/HCO$^{+}$ & 34 & 4.3 $\times$ 3.4 & $-$31.5 \\
         & CO            & 5  & 4.3 $\times$ 3.4 & $-$31.5 \\
Arp 193  & HCN/HCO$^{+}$ & 8  & 7.1 $\times$ 6.6 & $-$52.5 \\
NGC 1377 & HCN/HCO$^{+}$ & 23 & 7.1 $\times$ 4.0 & $-$13.6 \\
         & CO            & 4  & 6.7 $\times$ 3.6 & $-$18.8 \\
\enddata

\tablecomments{
Col.(1): Object name.
Col.(2): Observed line. HCN(1--0)/HCO$^{+}$(1--0) or CO(1--0).
Col.(3): Net on-source integration time in hours.
Col.(4): Beam size in arcsec $\times$ arcsec.
Col.(5): Position angle of the beam pattern.
It is 0$^{\circ}$ for the north-south direction, and increases 
counterclockwise on the sky plane.
}

\end{deluxetable}

\begin{deluxetable}{llclcccc}
\tabletypesize{\small}
\tablecaption{Observing log of infrared spectroscopy with IRTF SpeX 
\label{tab4}} 
\tablewidth{0pt}
\tablehead{
\colhead{Object} & 
\colhead{Date} & 
\colhead{Integration} & 
\multicolumn{4}{c}{Standard Stars} \\
\colhead{} & 
\colhead{(UT)} & 
\colhead{(Min)} &
\colhead{Name} &  
\colhead{$K$-mag} &  
\colhead{$L$-mag} &  
\colhead{Type} &  
\colhead{$T_{\rm eff}$ (K)}  \\
\colhead{(1)} & \colhead{(2)} & \colhead{(3)} & \colhead{(4)}
& \colhead{(5)} & \colhead{(6)}  & \colhead{(7)} & \colhead{(8)} 
}
\startdata
NGC 2623   & 2008 April 20 & 60 & HR 3262 & 3.9 & 3.9 & F6V & 6400 \\  
Mrk 266 SW & 2008 April 19 & 40 & HR 4767 & 4.9 & 4.8 & F9V--G0V & 6000 \\
Mrk 266 NE & 2008 April 19 & 60 & HR 4767 & 4.9 & 4.8 & F9V--G0V & 6000 \\
Arp 193    & 2008 April 20 & 60 & HR 4845 & 4.5 & 4.5 & G0V & 5930 \\  
\hline  
\enddata

\tablecomments{Column (1): Object name.
Col.(2): Observing date in UT.
Col.(3): Net on-source integration time in min.
Col.(4): Standard star name.
Col.(5): Adopted $K$-band magnitude.
Col.(6): Adopted $L$-band magnitude.
Col.(7): Stellar spectral type.
Col.(8): Effective temperature.
}

\end{deluxetable}

\begin{deluxetable}{lccccccc}
\tabletypesize{\scriptsize}
\tablecaption{Gaussian fitting parameters and nuclear fluxes of
HCN(1--0), HCO$^{+}$(1--0), and CO(1--0) emission lines \label{tab5}}   
\tablewidth{0pt}
\tablehead{
\colhead{Object} & \multicolumn{3}{c}{LSR velocity} &
\multicolumn{3}{c}{Flux} & \colhead{HCN(1--0)/HCO$^{+}$(1--0)} \\   
\colhead{} & \multicolumn{3}{c}{[km s$^{-1}$]} & 
\multicolumn{3}{c}{[Jy km s$^{-1}$]} & \colhead{Ratio} \\ 
\colhead{} & \colhead{HCN(1--0)} & \colhead{HCO$^{+}$(1--0)} &
\colhead{CO(1--0)} & \colhead{HCN(1--0)} & \colhead{HCO$^{+}$(1--0)}
& \colhead{CO(1--0)} & \colhead{} \\
\colhead{(1)} & \colhead{(2)} & \colhead{(3)} & \colhead{(4)} &
\colhead{(5)} & \colhead{(6)} & \colhead{(7)} & \colhead{(8)} 
}
\startdata
NGC 2623   & 5570 & 5620 & \nodata & 5.8  & 4.0 & 153 & 1.5 \\
Mrk 266 SW & 8290 & 8290 & 8370 & 1.0  & 3.5 & 88 & 0.3 \\
Mrk 266 NE & 8330 & 8370 & 8250 + 8501 & 1.1  & 0.8 & 18 & 1.4  \\
Arp 193    & 6905 + 7110 & 6850 + 7130 & \nodata & 10.8 & 12.7 & 202 & 0.9\\ 
NGC 1377   & 1500 & \nodata & 1740 & 2.2 & $<$1.2  & 9.6 & $>$1.8  \\
\enddata

\tablecomments{
Col.(1): Object name.
Cols.(2)--(4): LSR velocity \{v$_{\rm opt}$ $\equiv$ 
($\frac{\nu_0}{\nu}$ $-$ 1) $\times$ c\} of the
HCN(1--0), HCO$^{+}$(1--0), and CO(1--0) in [km s$^{-1}$].
Cols.(5)--(7): Flux of the HCN(1--0), HCO$^{+}$(1--0), 
and CO(1--0) in [Jy km s$^{-1}$] at the nuclear peak position.
For NGC 2623 and Arp 193, the CO(1--0) fluxes are adopted
from \citet{bry99}.
Col.(8): HCN(1--0)/HCO$^{+}$(1--0) ratio in brightness temperature 
($\propto$ $\lambda^{2}$ $\times$ flux density).
The ratio is not affected by possible absolute flux calibration
uncertainties in the NMA data (see $\S$5.1).
}

\end{deluxetable}

\begin{deluxetable}{llcl}
\tabletypesize{\small}
\tablecaption{Comparison of HCN(1--0), HCO$^{+}$(1--0), and CO(1--0) 
flux between our NMA interferometric measurements and single-dish
telescope's measurements in the literature \label{tab6}}   
\tablewidth{0pt}
\tablehead{
\colhead{Nucleus} & \colhead{Line} & \colhead{Flux} & \colhead{Reference} \\   
\colhead{(1)} & \colhead{(2)} & \colhead{(3)}  & \colhead{(4)} 
}
\startdata
Mrk 266  & CO(1--0)  & 180 (SW + NE) \tablenotemark{a} & This work \\ 
         &           & 206 \tablenotemark{b} & \citet{san86} \\ \hline
Arp 193  & HCN(1--0) & 10.8 & This work \\
         &           & 25 \tablenotemark{c} & \citet{sol92} \\
         &           & 6.3 \tablenotemark{d} & \citet{gra08} \\ 
         & HCO$^{+}$(1--0) & 12.7 & This work \\
         &           & 10.0 \tablenotemark{d} & \citet{gra08} \\ \hline
NGC 1377 & CO(1--0)  & 9.6 & This work \\ 
         &           & 47 \tablenotemark{e} & \citet{rou03} \\ 
\enddata

\tablecomments{
Col.(1): Object name.
Col.(2): HCN(1--0), HCO$^{+}$(1--0), or CO(1--0) line.
Col.(3): Flux in [Jy km s$^{-1}$].
Col.(4): Reference.
}
\tablenotetext{a}{For the CO(1--0) flux from the SW nucleus, we adopt
the total flux of 162 Jy km s$^{-1}$ ($\S$4.1), as
the spatially-extended component is very strong.}

\tablenotetext{b}{A conversion factor of 42 Jy K$^{-1}$ is assumed for
FCRAO 14-m telescope \citep{ken88}.
}

\tablenotetext{c}{A conversion factor of 4.4 Jy K$^{-1}$ is assumed for
IRAM 30-m telescope \citep{sol92}.
}

\tablenotetext{d}{A conversion factor of 6.0 Jy K$^{-1}$ (``K'' is
antenna temperature in this case) is assumed for IRAM 30-m telescope
\citep{gra08}. 
}

\tablenotetext{e}{A conversion factor of 27 Jy K$^{-1}$ is assumed for SEST
15-m telescope at 115 GHz
(http://www.ls.eso.org/lasilla/Telescopes/SEST/html/telescope-instruments/telescope/index.html).
}

\end{deluxetable}

\begin{deluxetable}{lcccccccc}
\tablecaption{Properties of hydrogen emission lines \label{tab7}}   
\tablewidth{0pt}
\tablehead{
\colhead{Object} & \multicolumn{2}{c}{Pa$\alpha$} & 
\multicolumn{2}{c}{Br$\alpha$} & 
\colhead{Br$\gamma$} & \colhead{H$_{2}$} & \colhead{Pa$\alpha$/Br$\alpha$} &
\colhead{H$_{2}$/Br$\gamma$} \\    
\colhead{} & \colhead{flux} & \colhead{FWHM} & \colhead{flux} &
\colhead{FWHM} & \colhead{flux} & \colhead{flux} & \colhead{} & \colhead{} \\
\colhead{(1)} & \colhead{(2)} & \colhead{(3)} & \colhead{(4)} &
\colhead{(5)} & \colhead{(6)} & \colhead{(7)} & \colhead{(8)} & 
\colhead{(9)} 
}
\startdata
NGC 2623   & 92 & 515 & 40      & 510     & 8  & 6.5 & 2.5     & 0.8 \\
Mrk 266 SW & 40 & 365 & \nodata & \nodata & 7  & 2.5 & \nodata & 0.3 \\ 
Mrk 266 NE & 50 & 510 & \nodata & \nodata & 2  & 7   & \nodata & 3.5 \\
Arp 193    & 72 & 440 & 39      & 465     & 11 & 7.5 & 2       & 0.7 \\ \hline
\enddata

\tablecomments{
Col.(1): Object name.
Col.(2): Pa$\alpha$ flux in [10$^{-15}$ ergs s$^{-1}$ cm$^{-2}$].
Col.(3): Pa$\alpha$ line width in FWHM in [km s$^{-1}$].
Col.(4): Br$\alpha$ flux in [10$^{-15}$ ergs s$^{-1}$ cm$^{-2}$].
Col.(5): Br$\alpha$ line width in FWHM in [km s$^{-1}$].
Col.(6): Br$\gamma$ flux in [10$^{-15}$ ergs s$^{-1}$ cm$^{-2}$].
Col.(7): H$_{2}$(1--0) S(1) flux in [10$^{-15}$ ergs s$^{-1}$ cm$^{-2}$].
Col.(8): Pa$\alpha$/Br$\alpha$ flux ratio.
Col.(9): H$_{2}$/Br$\gamma$ flux ratio.
}

\end{deluxetable}

\begin{deluxetable}{lcccc}
\tablecaption{Properties of 3.3-$\mu$m PAH emission and 2.3-$\mu$m CO
absorption \label{tab8}}    
\tablewidth{0pt}
\tablehead{
\colhead{Object} & \colhead{f$_{3.3 \rm PAH}$} & 
\colhead{L$_{3.3 \rm PAH}$} & \colhead{EW$_{\rm 3.3PAH}$} &
\colhead{CO$_{\rm spec}$} \\    
\colhead{(1)} & \colhead{(2)} & \colhead{(3)} & \colhead{(4)} & \colhead{(5)} 
}
\startdata
NGC 2623   & 16 & 1.0 & 125 & 0.27 \\
Mrk 266 SW & 9  & 1.4 & 70  & 0.19 \\ 
Mrk 266 NE & 3  & 0.4 & 30  & 0.24 \\
Arp 193    & 21 & 2.2 & 140 & 0.21 \\ \hline
\enddata

\tablecomments{
Col.(1): Object name.
Col.(2): Observed flux of 3.3-$\mu$m PAH emission in 
[10$^{-14}$ ergs s$^{-1}$ cm$^{-2}$].
Col.(3): Observed luminosity of 3.3-$\mu$m PAH emission 
in [10$^{41}$ ergs s$^{-1}$].
Col.(4): Rest-frame equivalent width of the 3.3-$\mu$m PAH emission 
in [nm]. 
Col.(5): Spectroscopic CO index defined by \citet{doy94}.
}

\end{deluxetable}

\clearpage

\begin{figure}
\includegraphics[angle=0,scale=.4]{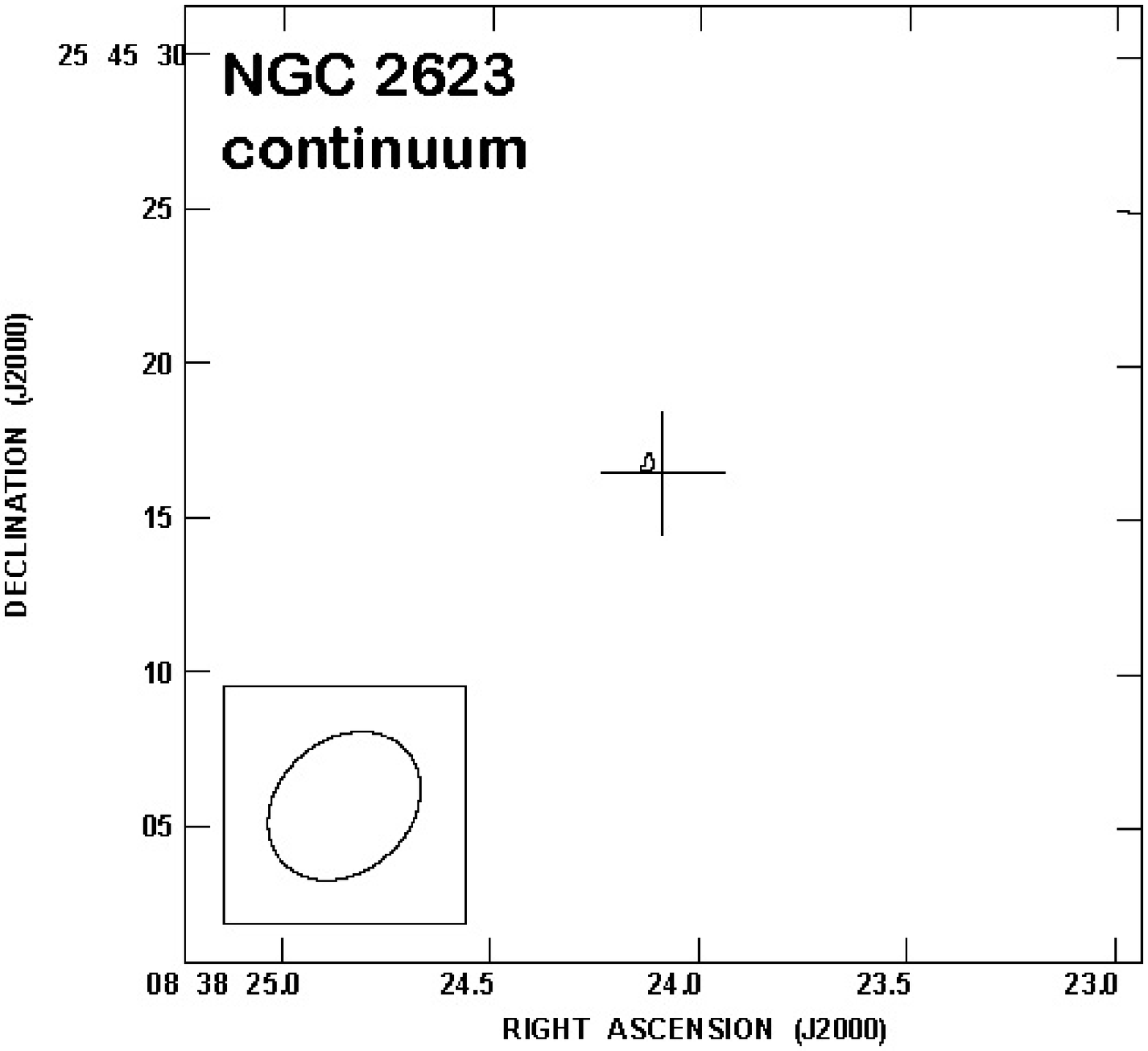} \hspace{1.3cm}
\includegraphics[angle=0,scale=.4]{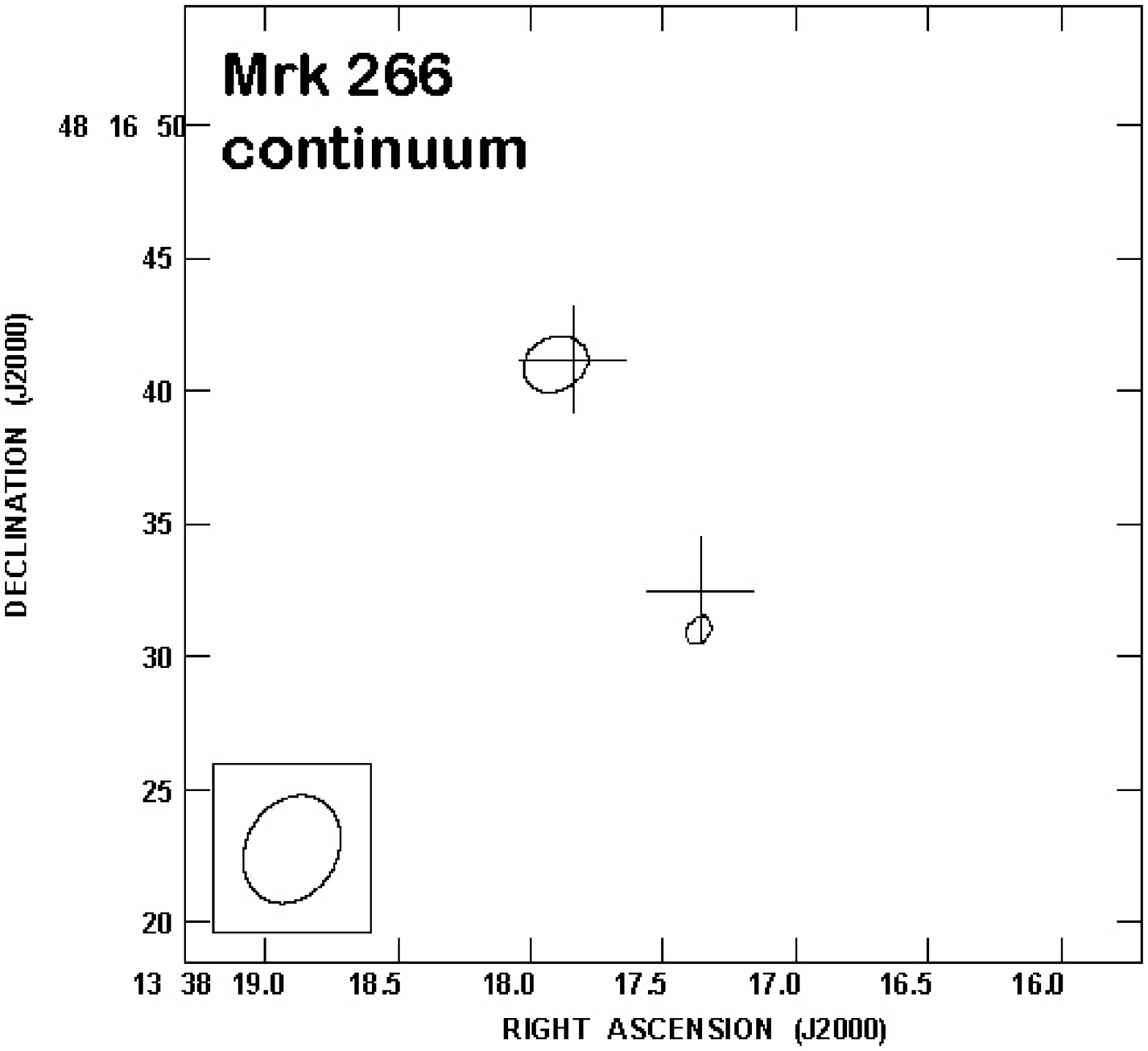} \\
\includegraphics[angle=0,scale=.4]{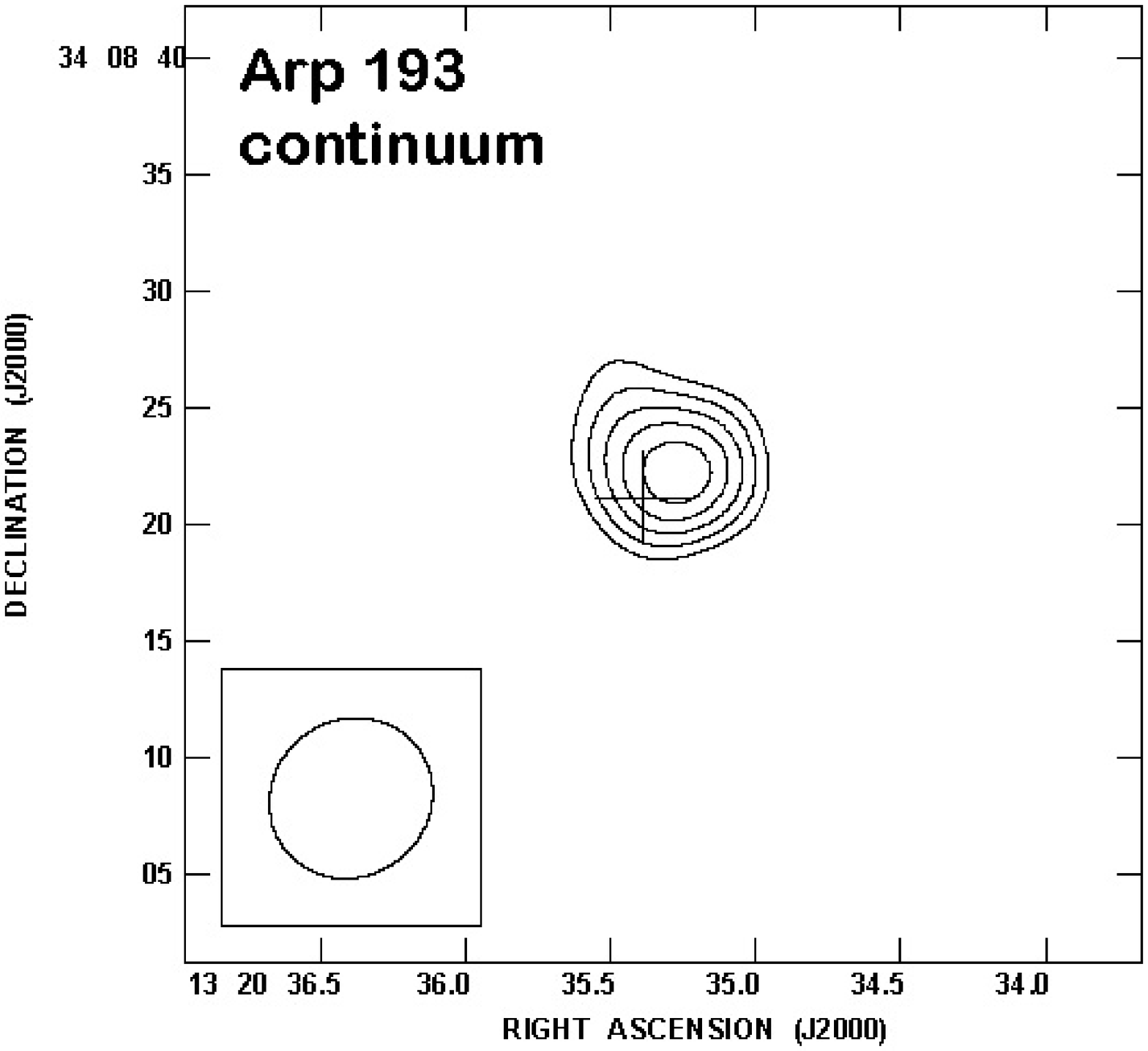} \\
\caption{
Continuum maps of NGC 2623, Mrk 266, and Arp 193 at $\nu$ $\sim$ 86--89
GHz. 
The crosses show the coordinates of main nuclei.
The coordinates in J2000 are
(08$^{h}$38$^{m}$24.09$^{s}$, $+$25$^{\circ}$45$'$16$\farcs$5) for 
NGC 2623,
(13$^{h}$38$^{m}$17.36$^{s}$, $+$48$^{\circ}$16$'$32$\farcs$5) for
Mrk 266 SW,
(13$^{h}$38$^{m}$17.84$^{s}$, $+$48$^{\circ}$16$'$41$\farcs$1) for
Mrk 266 NE, and 
(13$^{h}$20$^{m}$35.39$^{s}$, $+$34$^{\circ}$08$'$21$\farcs$1) for
Arp 193.
The nuclear coordinates of NGC 2623 and Arp 193 are adopted from
interferometric CO(1--0) maps by \citet{bry99}, after conversion from
B1950 to J2000, using NED. 
That of Mrk 266 is estimated from {\it Chandra} X-ray data 
by \citet{bra07}.
The contours are 1.3 $\times$ 3 mJy beam$^{-1}$ for NGC 2623, 
0.8 $\times$ 3 mJy beam$^{-1}$ for Mrk 266, and 
1.5 $\times$ (4, 5, 6, 7, 8) mJy beam$^{-1}$ for Arp 193.
The continuum fluxes at the peak positions are 
$\sim$4 mJy, $\sim$2.8 mJy, $\sim$2.3 mJy, and $\sim$15 mJy for 
NGC 2623, Mrk 266NE, Mrk 266SW, and Arp 193, respectively. 
Beam patterns are shown in the small squares at the lower-left corners. 
}
\end{figure}

\begin{figure}
\includegraphics[angle=0,scale=.4]{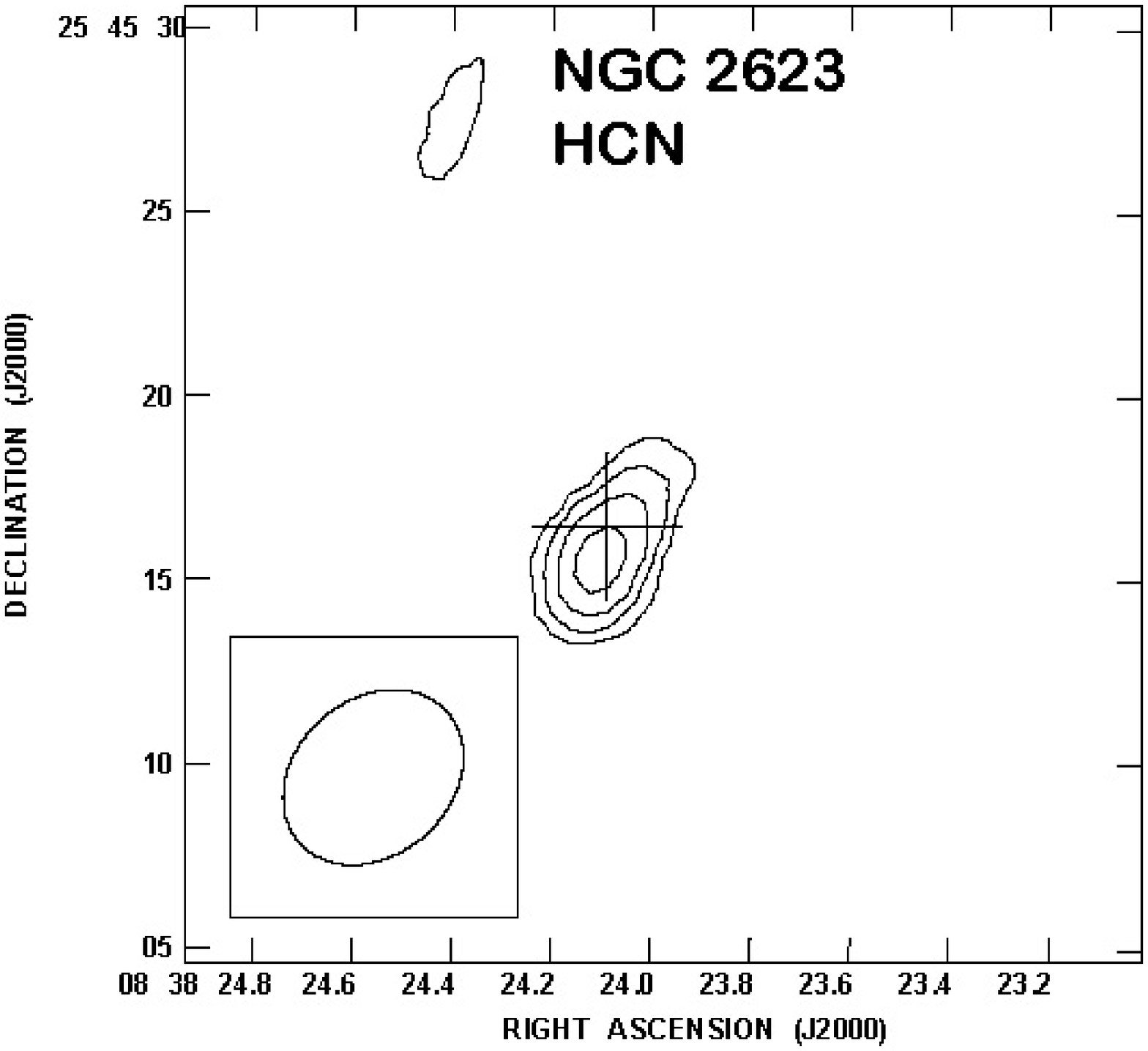} \hspace{0.8cm} 
\includegraphics[angle=0,scale=.4]{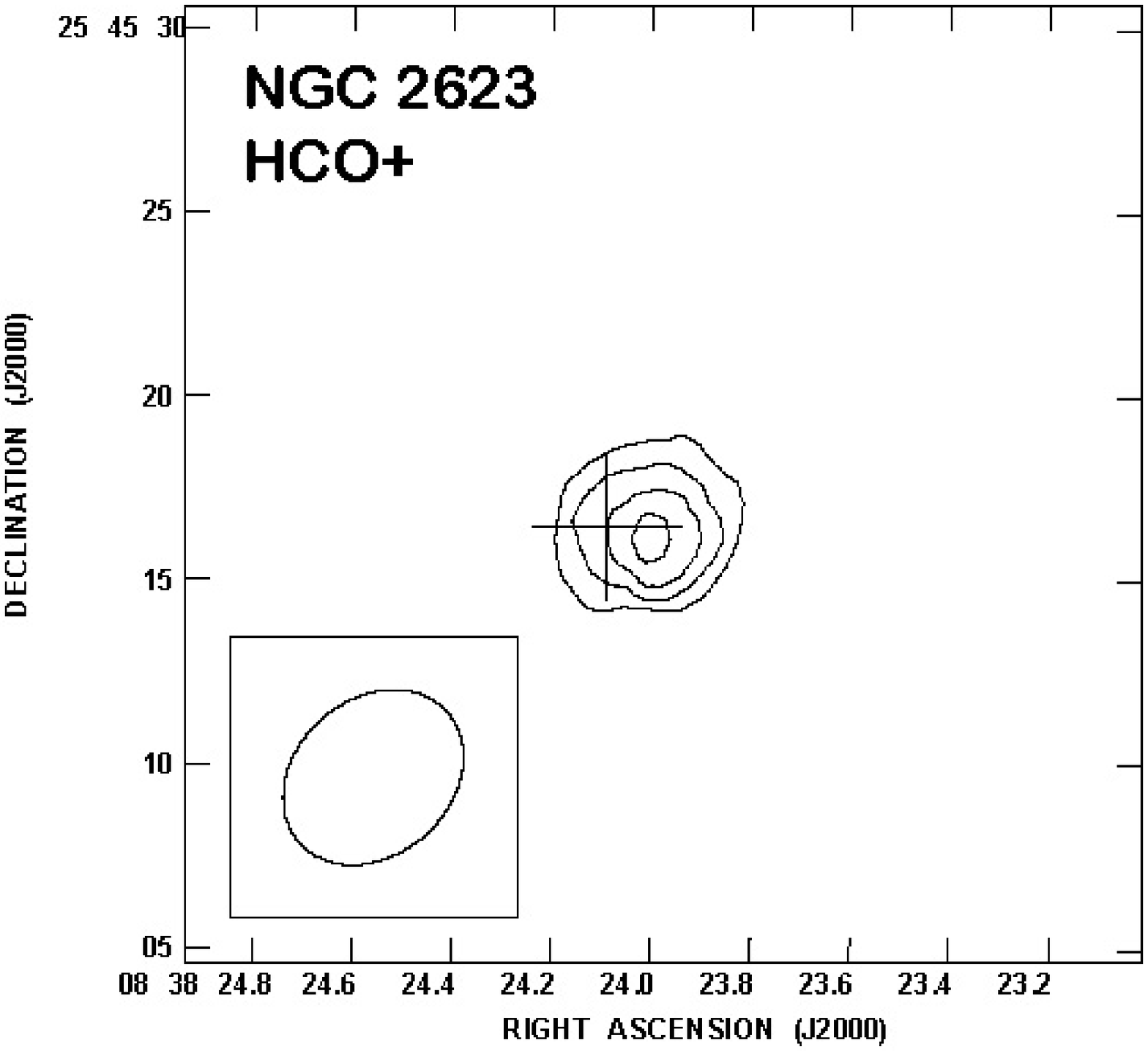} \\
\includegraphics[angle=0,scale=.4]{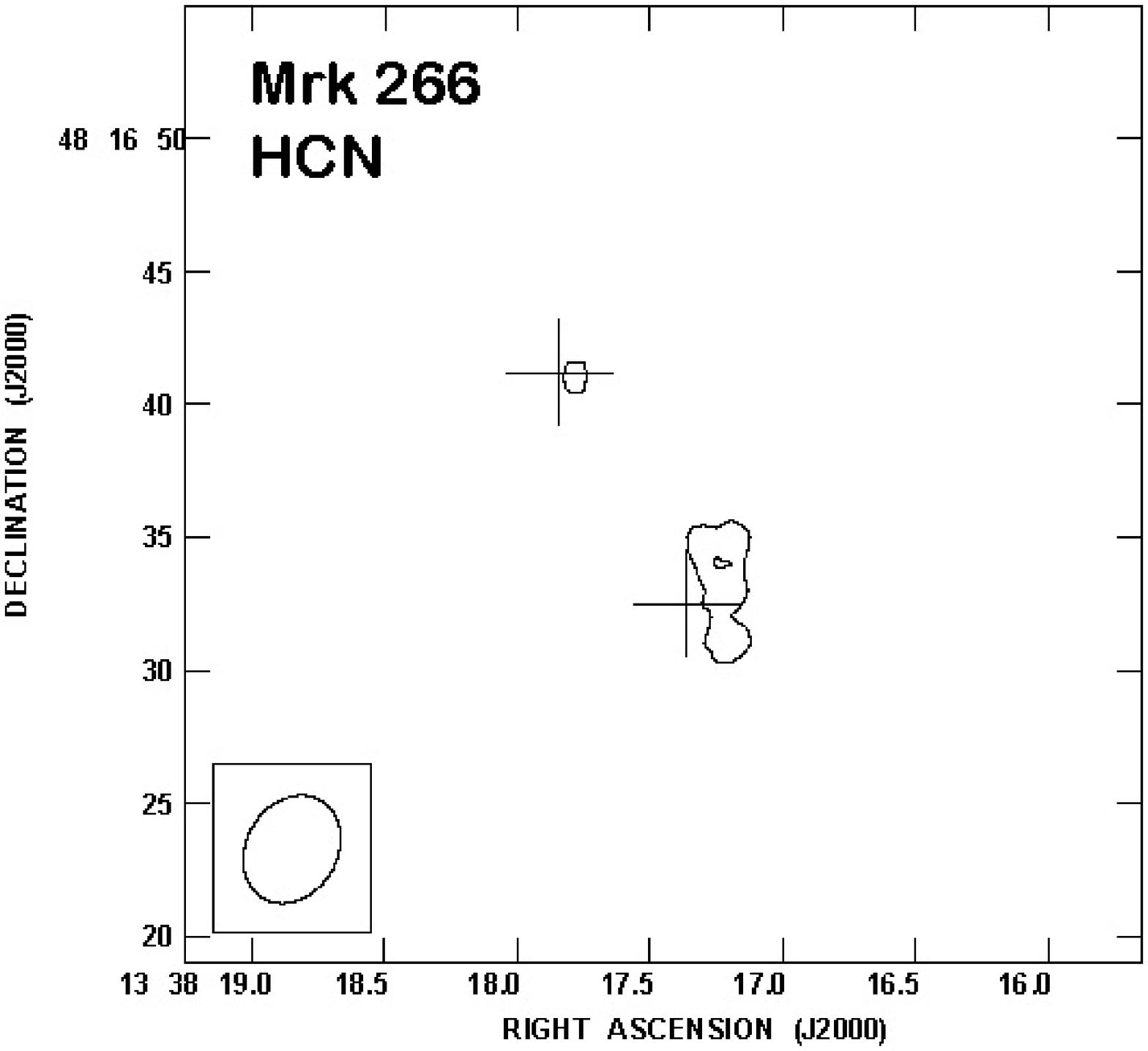} \hspace{0.8cm} 
\includegraphics[angle=0,scale=.4]{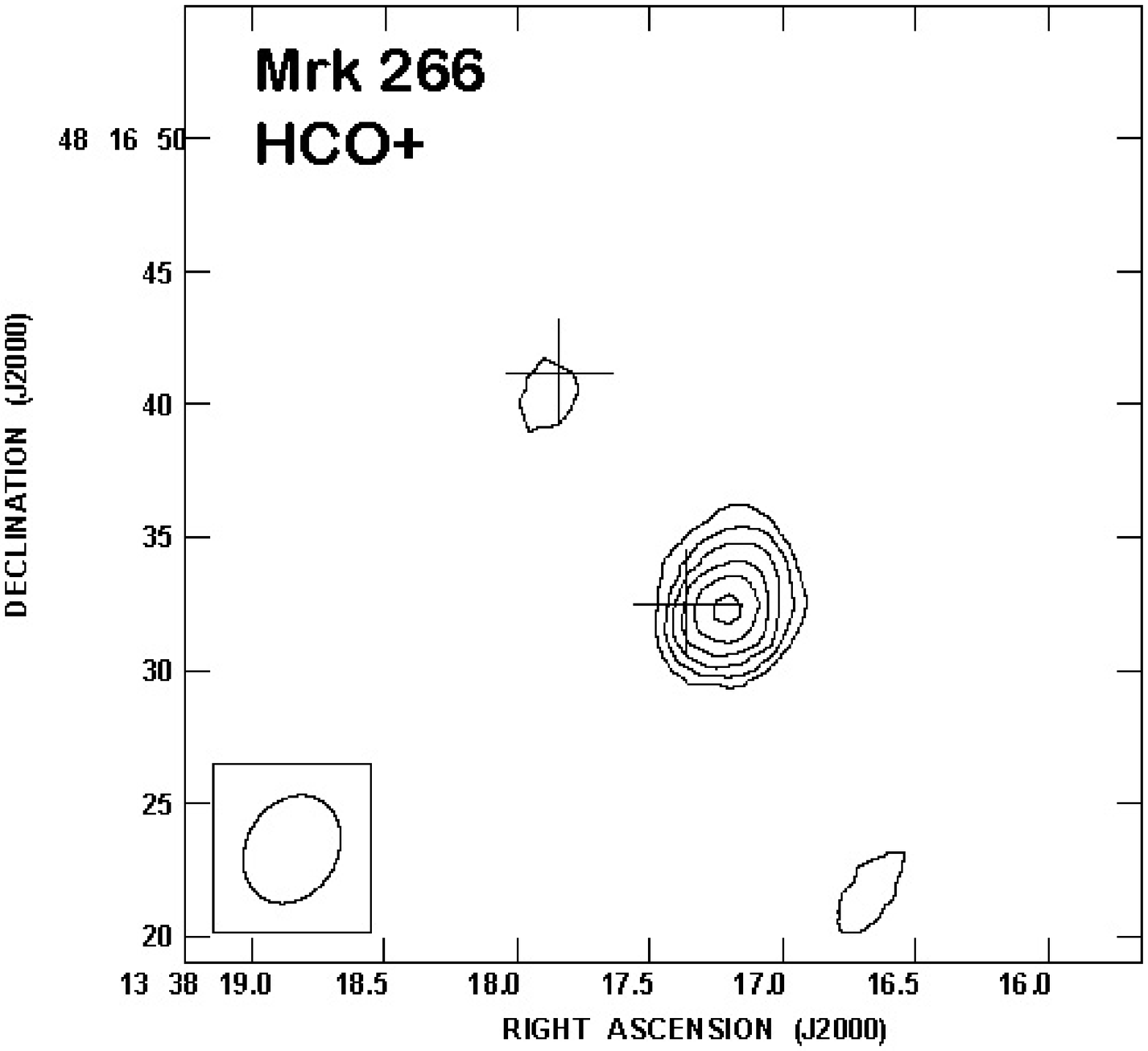} \\
\includegraphics[angle=0,scale=.4]{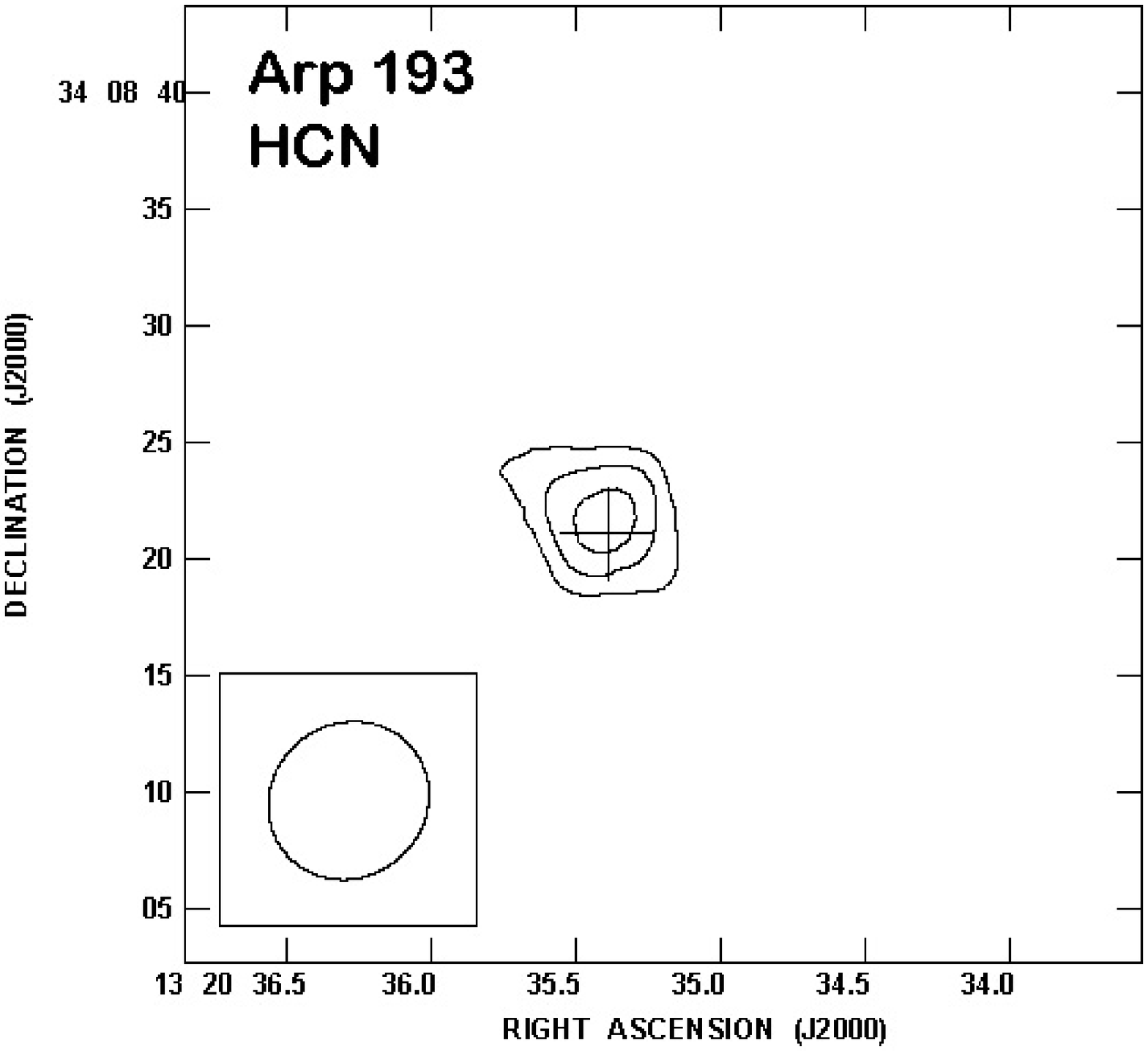} \hspace{0.8cm} 
\includegraphics[angle=0,scale=.4]{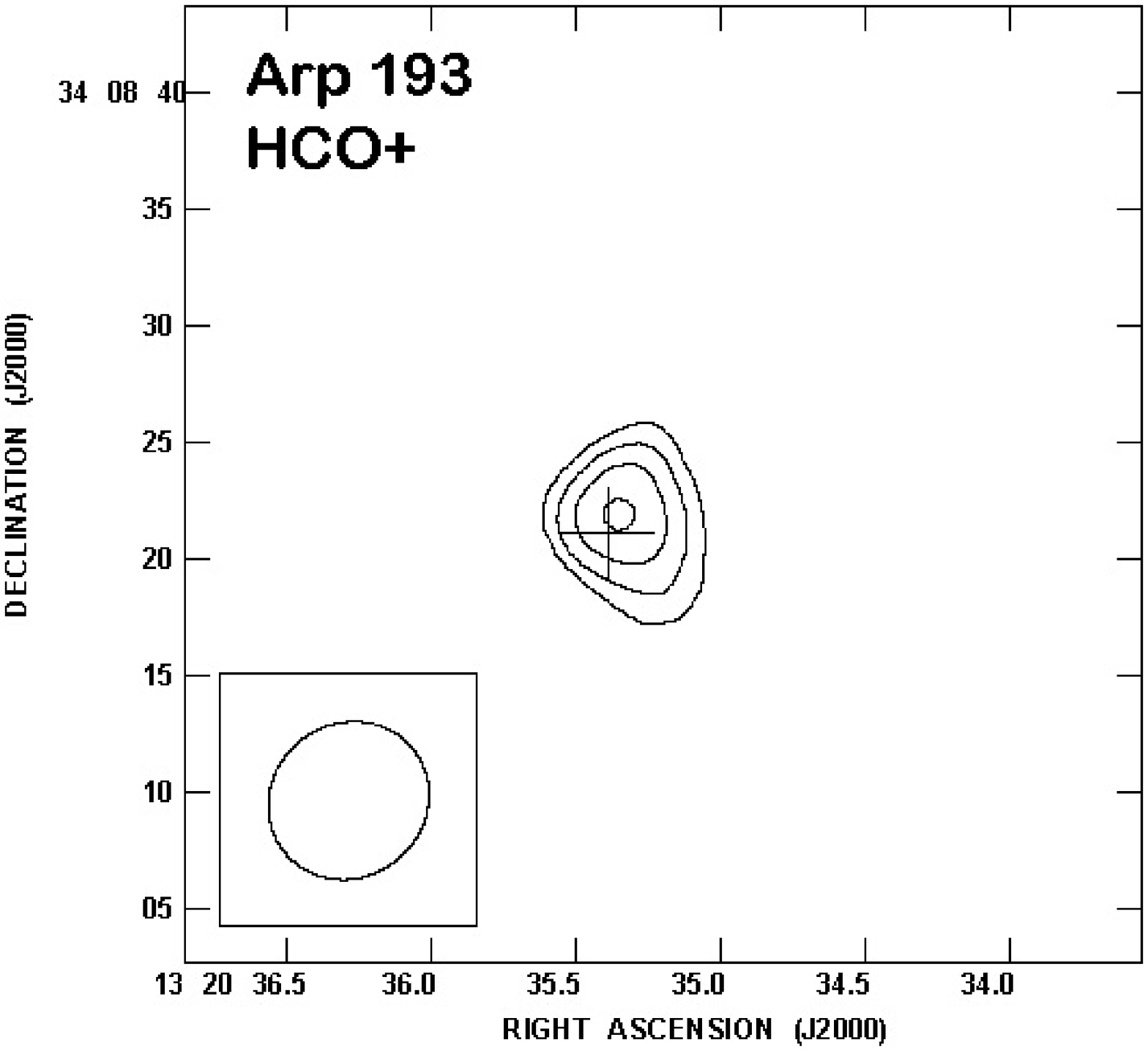} \\
\end{figure}

\clearpage

\begin{figure}
\includegraphics[angle=0,scale=.4]{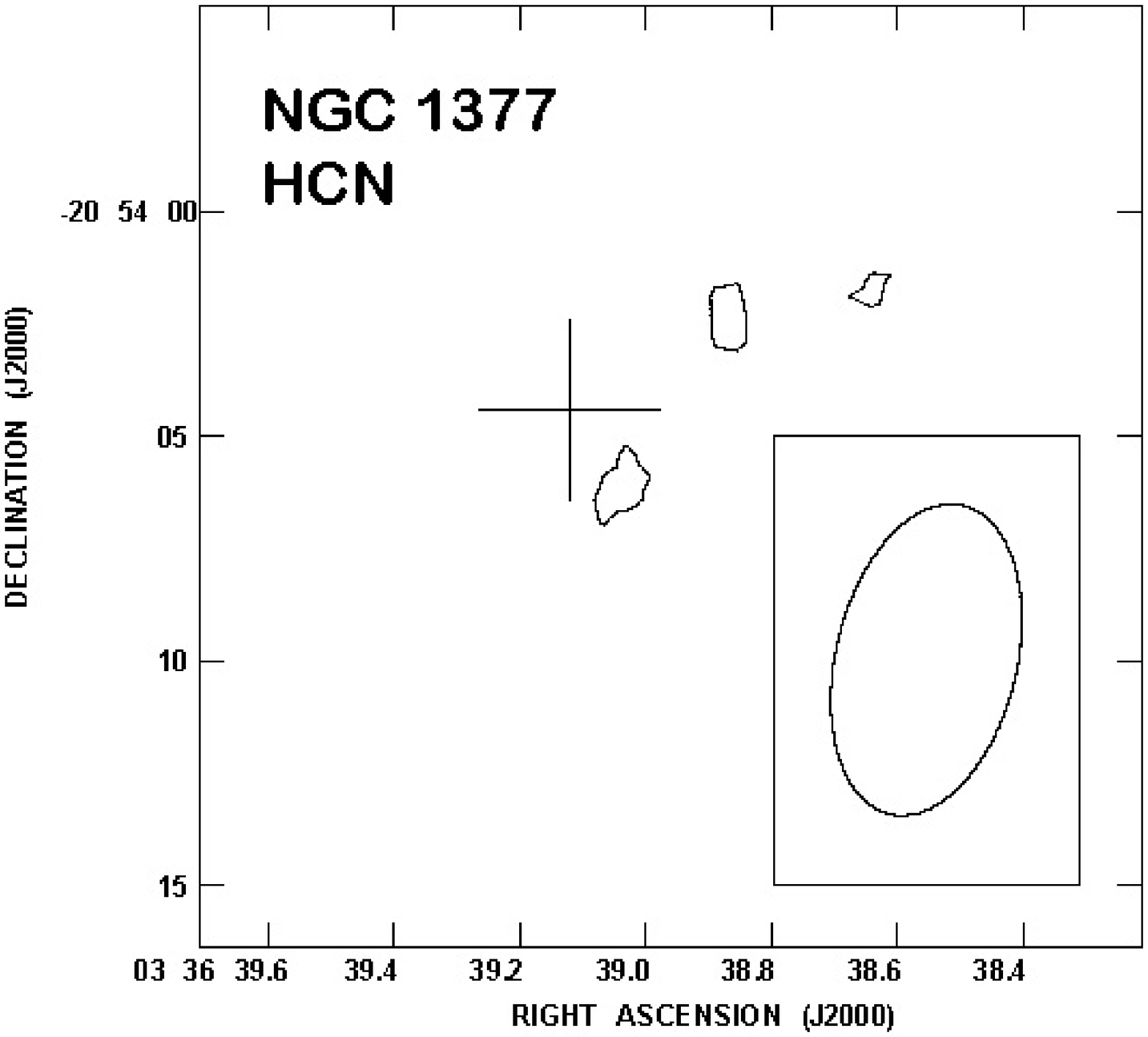} \hspace{0.8cm}
\includegraphics[angle=0,scale=.4]{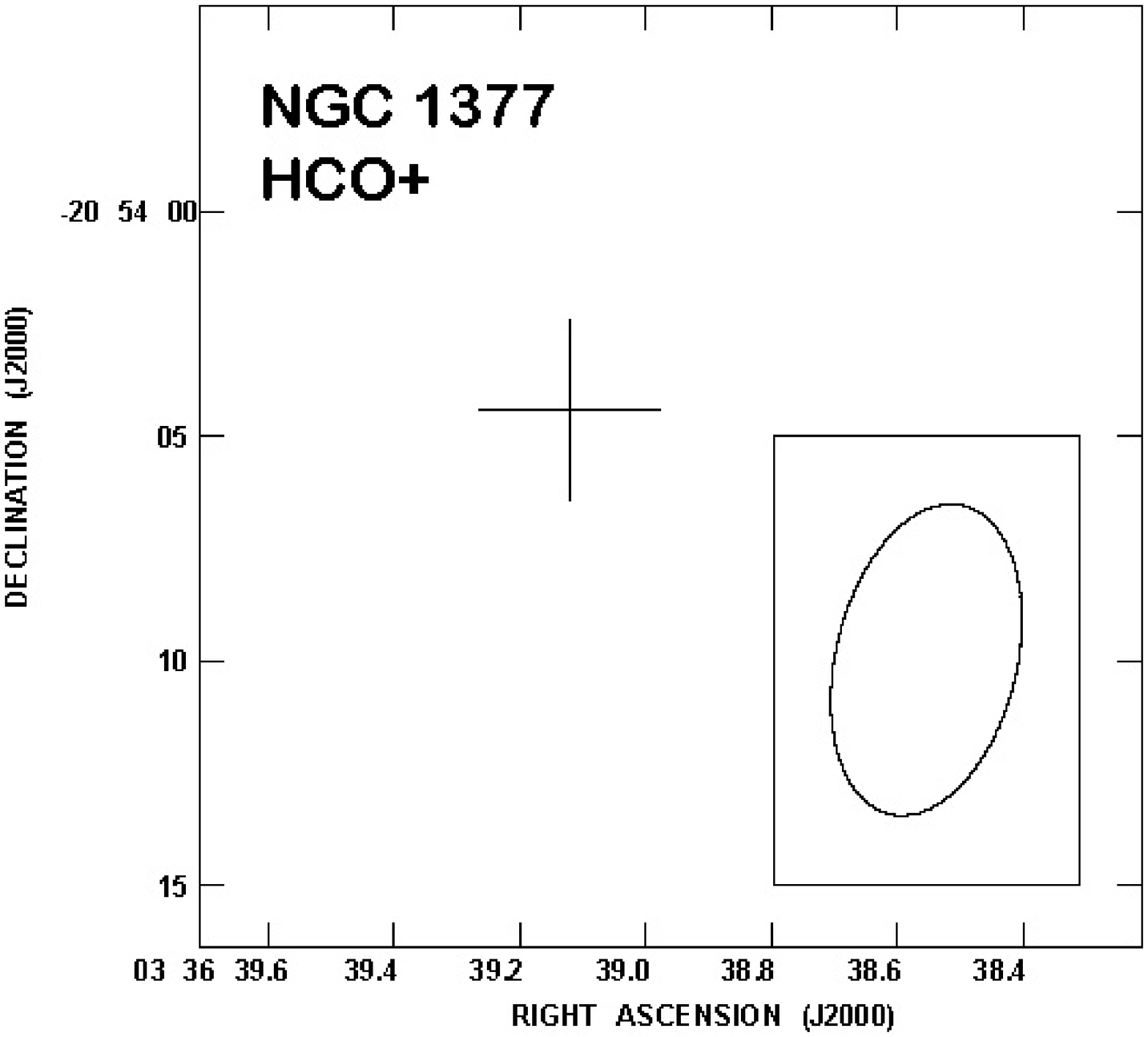} \\
\caption{
{\it Left}: HCN(1--0) emission map.
{\it Right}: HCO$^{+}$(1--0) emission map.
The crosses show the coordinates of main nuclei.
For NGC 2623, Mrk 266, and Arp 193, the same coordinates as used in
Figure 1 are shown.   
For NGC 1377, the coordinate in J2000 is 
(03$^{h}$36$^{m}$39.12$^{s}$, $-$20$^{\circ}$54$'$04$\farcs$4), 
estimated from optical $B$-band data by \citet{rou06}.  
Continuum emission is subtracted for NGC 2623 and Arp 193, 
but not for Mrk 266 and NGC 1377 because continuum levels are
different for the NE and SW nuclei for Mrk 266 and 
continuum emission is not detected for NGC 1377 ($<$2 mJy).
For NGC 2623, the contours are 0.66 $\times$ (4, 5, 6, 7) Jy km s$^{-1}$ 
for HCN, and 0.73 $\times$ (4, 5, 6, 7) Jy km s$^{-1}$ for HCO$^{+}$.
For Mrk 266, the contours are 0.62 $\times$ (4, 5) Jy km s$^{-1}$ 
for HCN, and 0.55 $\times$ (3, 4, 5, 6, 7, 8) Jy km s$^{-1}$  for HCO$^{+}$.
For Arp 193, the contours are 1.0 $\times$ (6, 8, 10) Jy km s$^{-1}$ 
for HCN, and 1.0 $\times$ (6, 8, 10, 12) Jy km s$^{-1}$  for HCO$^{+}$.
For NGC 1317, the contours are 0.5 $\times$ 4 Jy km s$^{-1}$  for HCN,
and 0.5 $\times$ 3 Jy km s$^{-1}$ for HCO$^{+}$. 
}
\end{figure}

\begin{figure}
\includegraphics[angle=0,scale=.45]{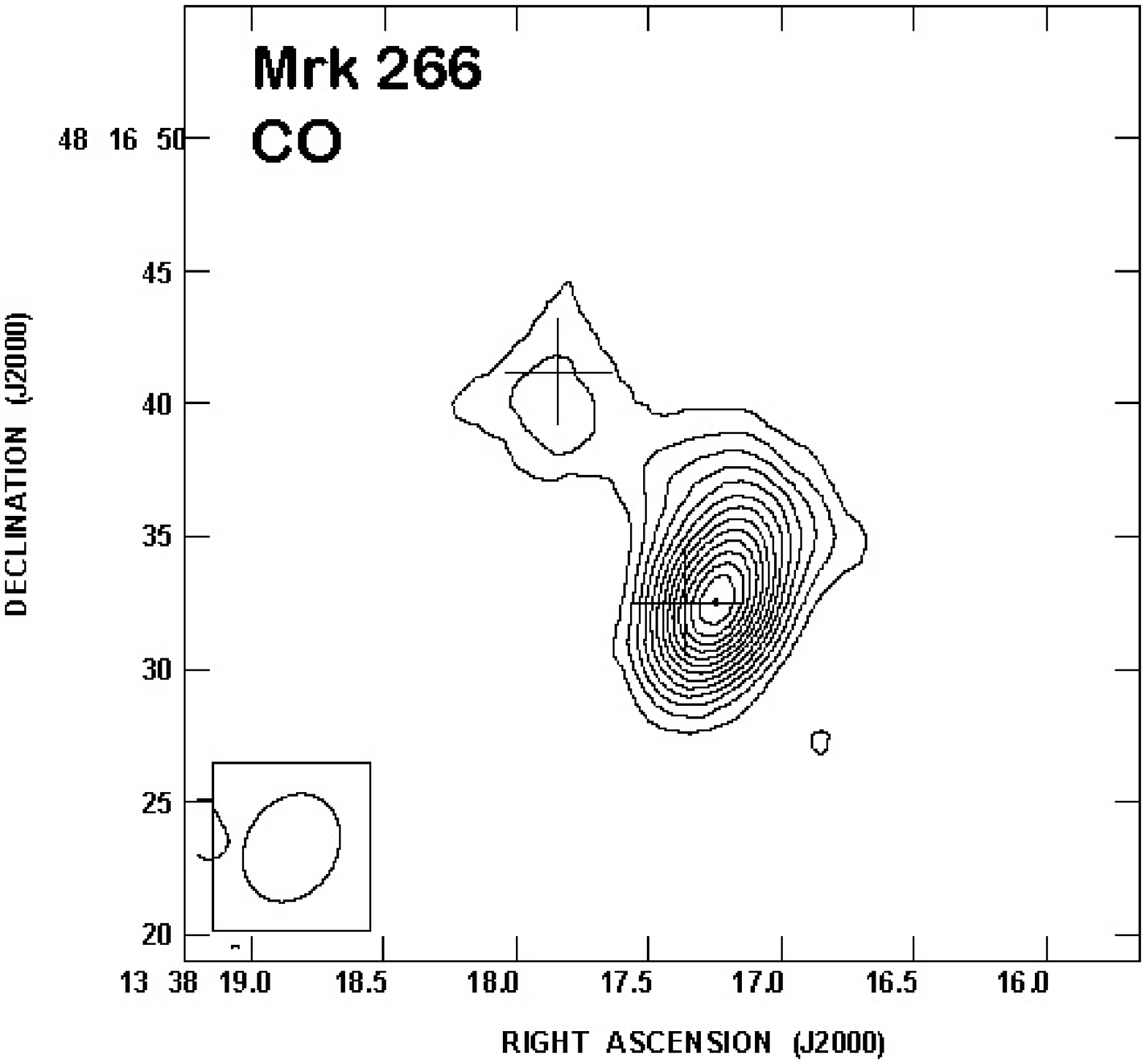} \hspace{0.8cm}
\includegraphics[angle=0,scale=.45]{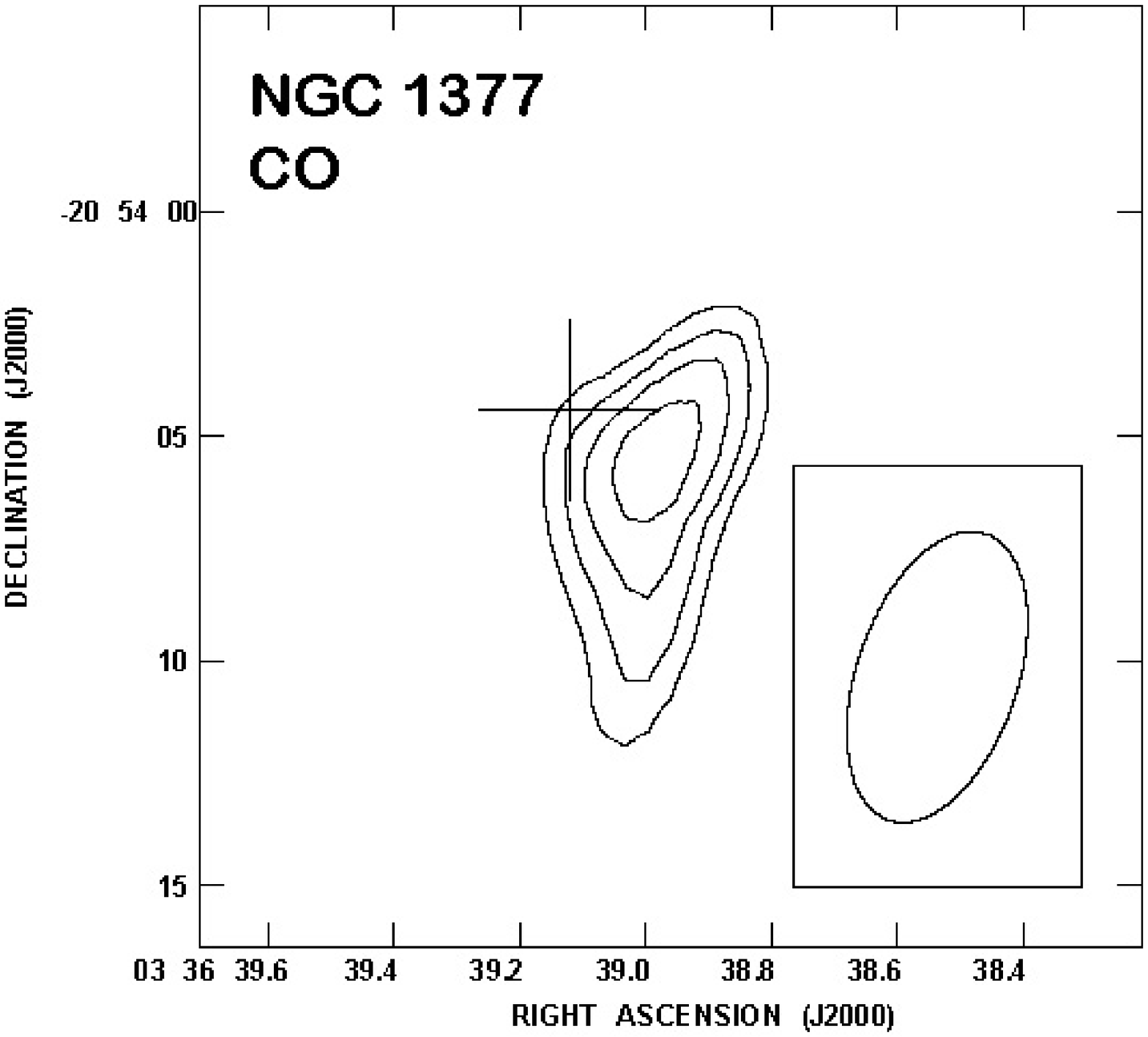} \\
\caption{
CO(1--0) emission for Mrk 266 and NGC 1377.
No continuum subtraction was attempted, because the continuum emission
is undetected ($<$5 mJy and $<$6 mJy for Mrk 266 and NGC 1377,
respectively) and negligible compared to the strong CO(1--0) emission.
The contours are 2.85 $\times$ (3, 5, 7, 9, 11, 13, 15, 17, 19, 21, 23, 25,
27, 29) Jy km s$^{-1}$ for Mrk 266, and 1.21 $\times$ (5, 6, 7, 8) 
Jy km s$^{-1}$ for NGC 1377. 
}
\end{figure}

\begin{figure}
\begin{center}
\includegraphics[angle=0,scale=0.85]{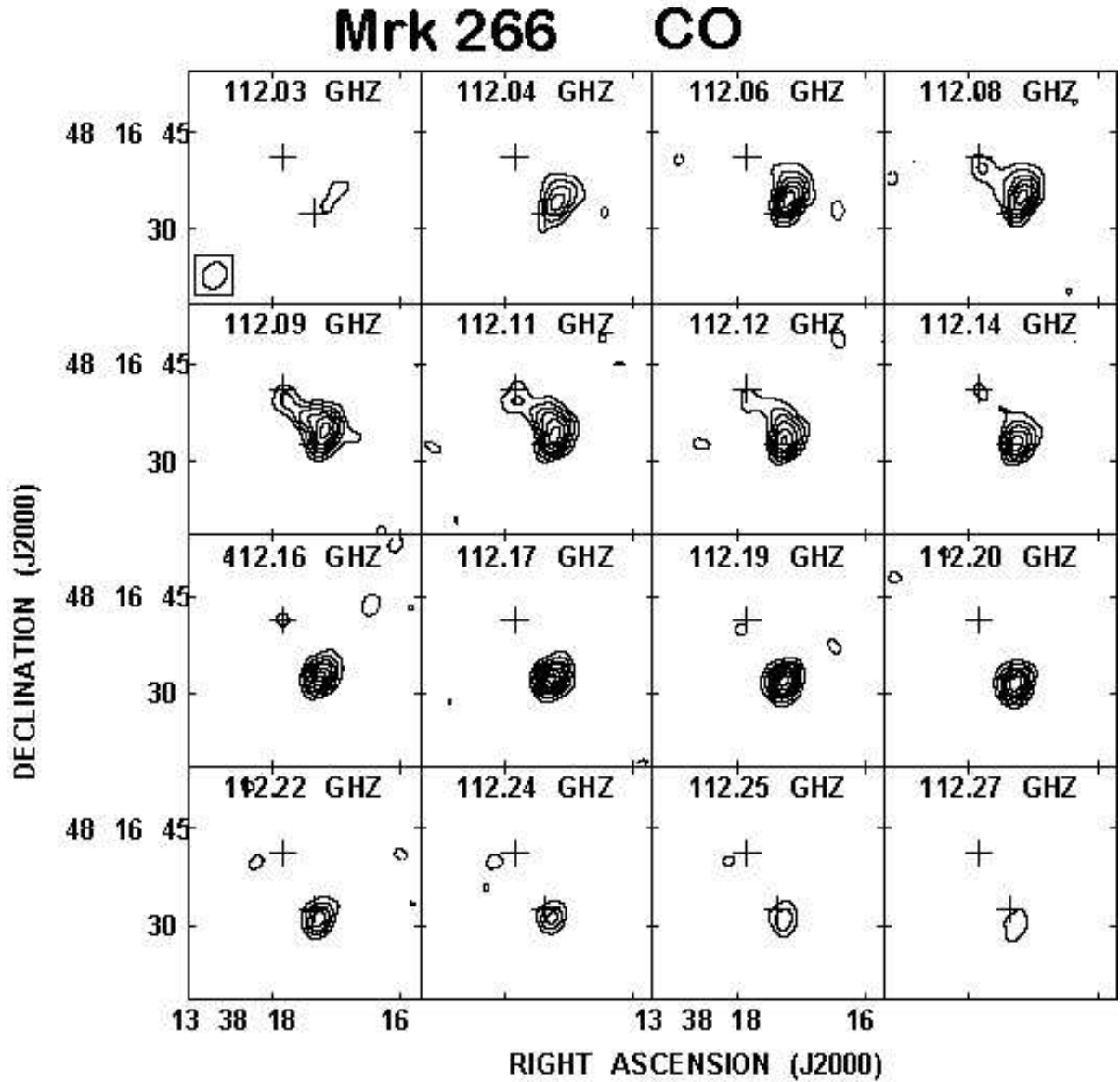} \\
\end{center}
\caption{
Channel map of CO(1--0) emission for Mrk 266. 
The contours are 12 $\times$ ($-$3, 3, 5, 7, 9, 11, 13, 15)
mJy beam$^{-1}$.
The rms noise level is $\sim$12 mJy beam$^{-1}$.
}
\end{figure}

\begin{figure}
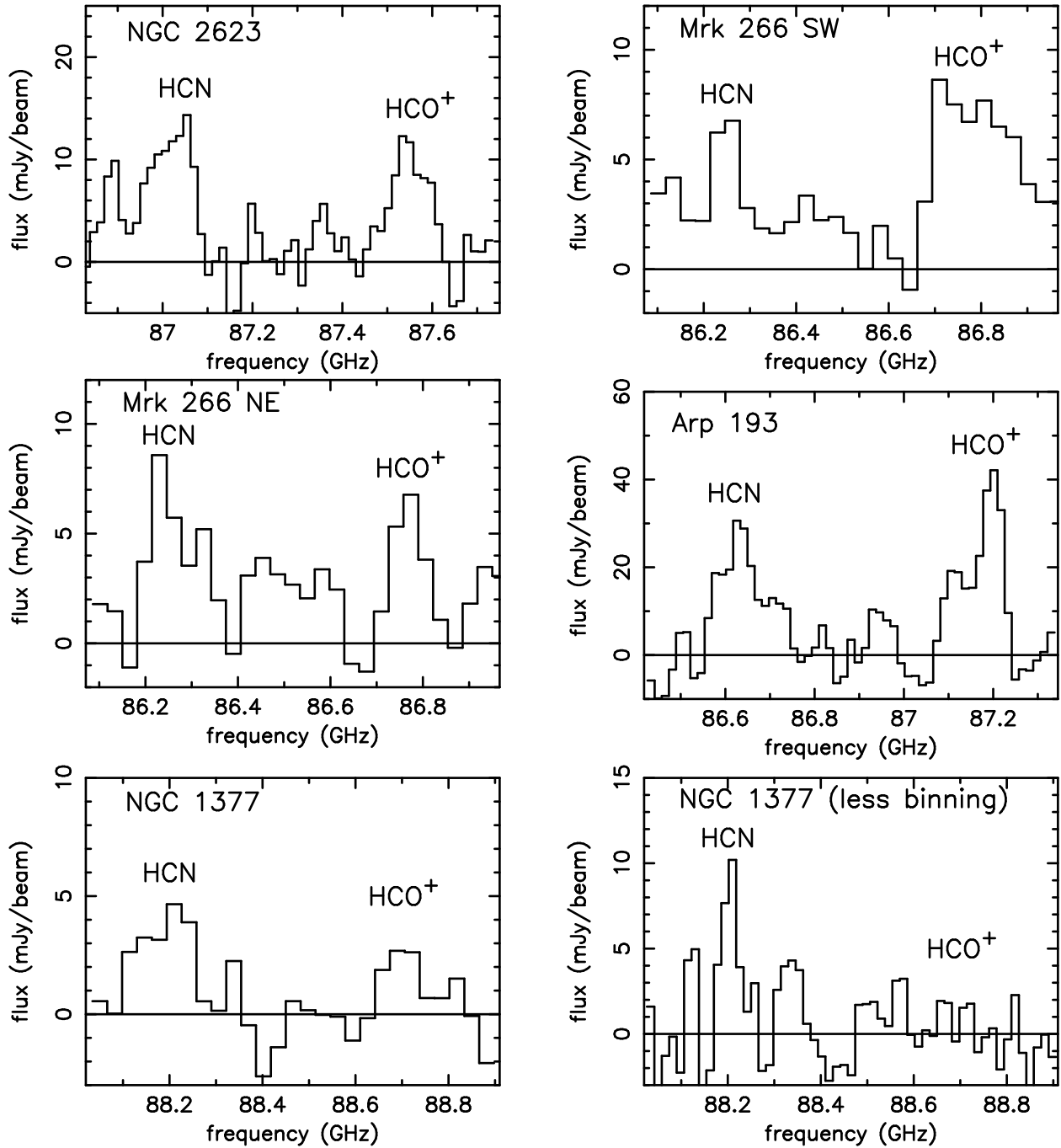

\includegraphics[angle=-90,scale=.35]{f5a.eps} \hspace{0.8cm} 
\includegraphics[angle=-90,scale=.35]{f5b.eps} \\
\includegraphics[angle=-90,scale=.35]{f5c.eps} \hspace{0.8cm} 
\includegraphics[angle=-90,scale=.35]{f5d.eps} \\
\includegraphics[angle=-90,scale=.35]{f5e.eps} \hspace{0.8cm} 
\includegraphics[angle=-90,scale=.35]{f5f.eps} \\
\caption{
HCN(1--0) and HCO$^{+}$(1--0) spectra of observed LIRGs.
The abscissa is the observed frequency in GHz and the ordinate is flux in
mJy beam$^{-1}$. 
For NGC 2623 and Arp 193, continuum emission is subtracted, but
not for Mrk 266 and NGC 1377.
The rms noise levels per spectral bin in the final spectra are 
$\sim$4 mJy (NGC 2623), $\sim$2.3 mJy (Mrk 266), $\sim$7 mJy (Arp 193),
$\sim$1.8 mJy (NGC 1377), and $\sim$2.6 mJy (NGC 1377, less binning). 
The horizontal solid line indicates the zero flux level. 
For the two spectra of NGC 1377, a clean map is first created, and then
a spectrum is extracted at the HCN(1--0) peak position in each map. 
}
\end{figure}

\begin{figure}
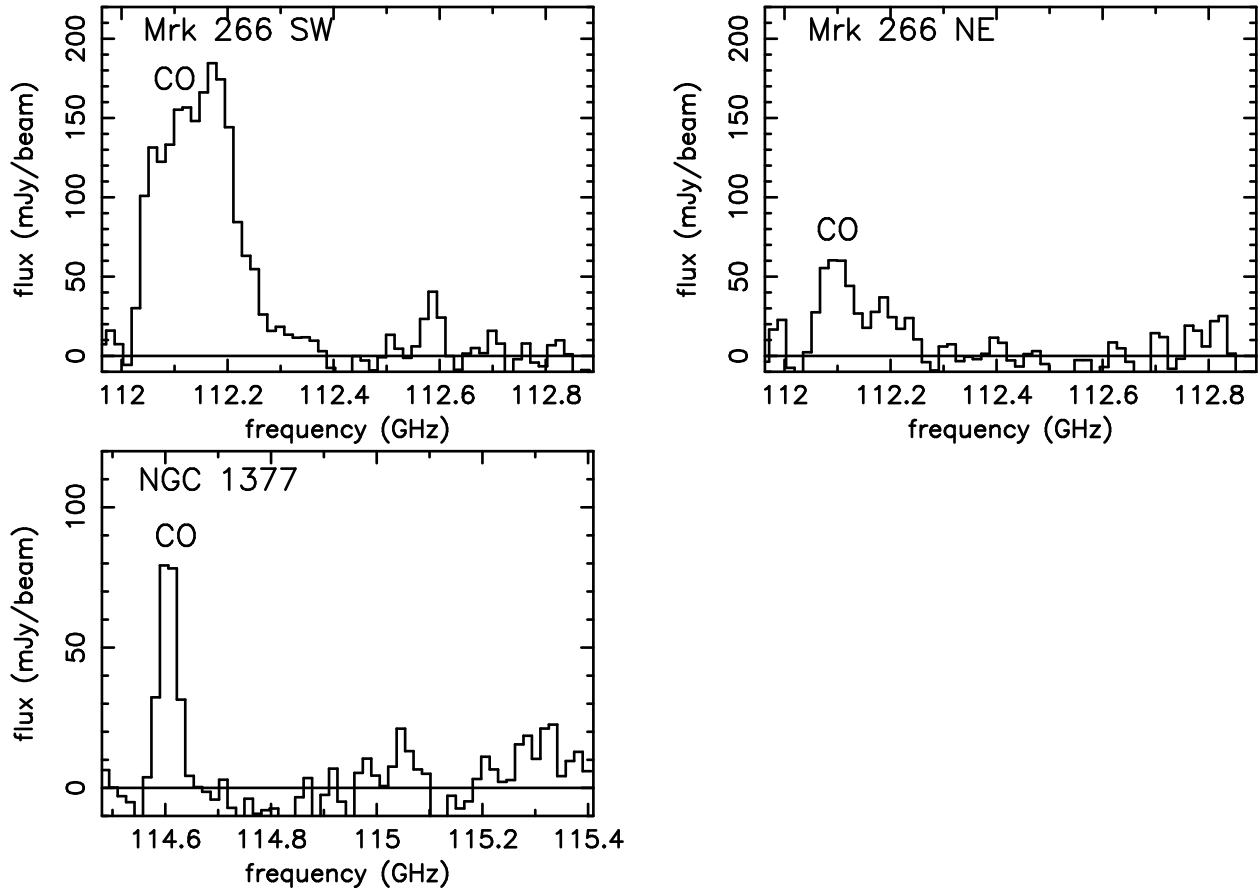

\includegraphics[angle=-90,scale=.35]{f6a.eps} \hspace{0.8cm} 
\includegraphics[angle=-90,scale=.35]{f6b.eps} \\
\includegraphics[angle=-90,scale=.35]{f6c.eps} \hspace{0.8cm} 
\caption{
CO(1--0) spectra of Mrk 266 and NGC 1377.
The abscissa is the observed frequency in GHz and the ordinate is flux
in mJy beam$^{-1}$. 
No continuum subtraction is attempted. 
The rms noise levels per spectral bin in the final spectra are 
$\sim$12 mJy and $\sim$9 mJy for Mrk 266 and NGC 1377, respectively.
The horizontal solid line marks the zero flux level. 
}
\end{figure}

\begin{figure}
\includegraphics[angle=-90,scale=.35]{f7a.eps} \hspace{0.8cm}
\includegraphics[angle=-90,scale=.35]{f7b.eps} \\
\includegraphics[angle=-90,scale=.35]{f7c.eps} \hspace{0.8cm}
\includegraphics[angle=-90,scale=.35]{f7d.eps} \\
\includegraphics[angle=-90,scale=.35]{f7e.eps} \hspace{0.8cm} 
\includegraphics[angle=-90,scale=.35]{f7f.eps} \\
\includegraphics[angle=-90,scale=.35]{f7g.eps} \hspace{0.8cm} 
\includegraphics[angle=-90,scale=.35]{f7h.eps} \\
\end{figure}

\clearpage

\begin{figure}
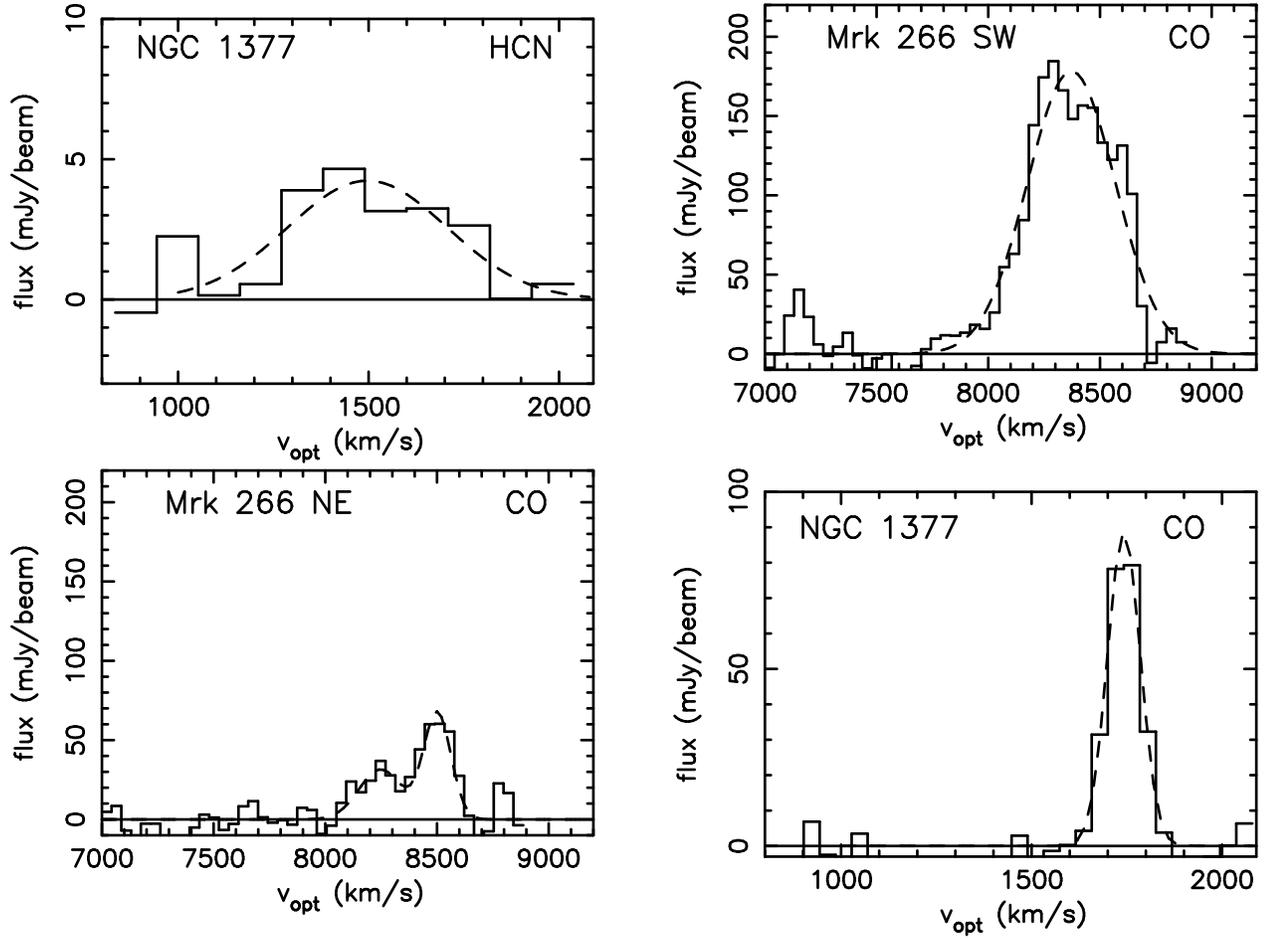

\includegraphics[angle=-90,scale=.35]{f7i.eps} \hspace{0.8cm}
\includegraphics[angle=-90,scale=.35]{f7j.eps} \\
\includegraphics[angle=-90,scale=.35]{f7k.eps} \hspace{0.8cm}
\includegraphics[angle=-90,scale=.35]{f7l.eps} \\
\caption{
Gaussian fits to the detected HCN(1--0), HCO$^{+}$(1--0), and CO(1--0) 
emission lines.  The abscissa is the LSR velocity \{v$_{\rm opt}$ $\equiv$
($\frac{\nu_0}{\nu}$ $-$ 1) $\times$ c\} in km s$^{-1}$ and the
ordinate is flux in mJy beam$^{-1}$.  Although single Gaussian fits 
are used by default, two Gaussian fits are attempted for double-peaked lines.
These lines include HCN(1--0) and HCO$^{+}$(1--0) of Arp 193, and
CO(1--0) of Mrk 266 NE. 
For Arp 193 HCN(1--0), even though the double-peaked signature is not 
clear, we fit with two Gaussians because the HCN(1--0) profile measured
with the IRAM 30-m single-dish telescope \citep{gra08} is double-peaked. 
For Mrk 266 and NGC 1377, a constant continuum is assumed and is set as
a free parameter because continuum emission was not subtracted.
For other sources, the continuum level is set as zero. 
The adopted continuum levels are shown as horizontal solid straight
lines for all sources. 
}
\end{figure}

\begin{figure}
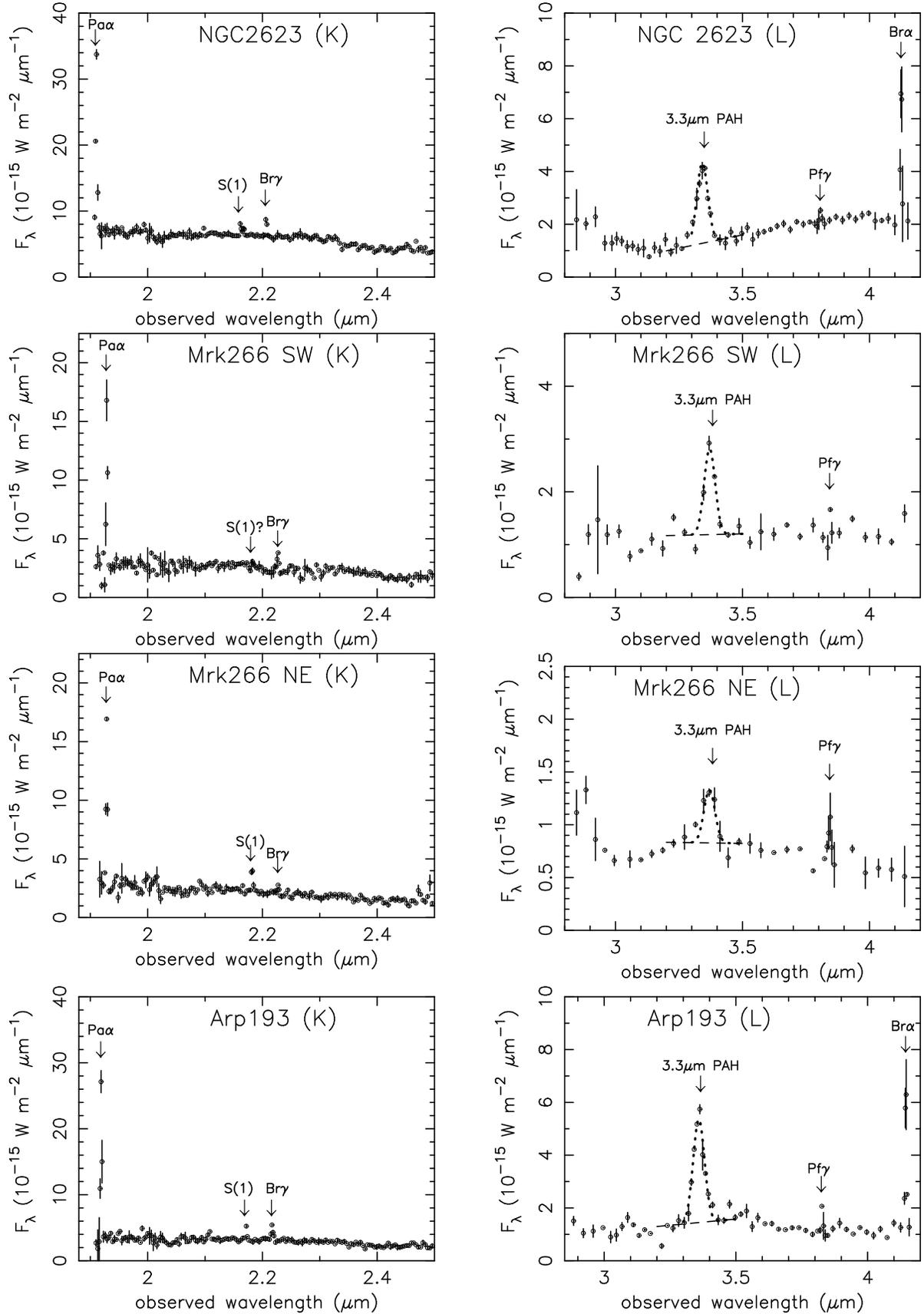

\begin{center}
\includegraphics[angle=-90,scale=.33]{f8a.eps} \hspace{0.8cm}
\includegraphics[angle=-90,scale=.33]{f8b.eps} \\
\includegraphics[angle=-90,scale=.33]{f8c.eps} \hspace{0.8cm}
\includegraphics[angle=-90,scale=.33]{f8d.eps} \\
\includegraphics[angle=-90,scale=.33]{f8e.eps} \hspace{0.8cm}
\includegraphics[angle=-90,scale=.33]{f8f.eps} \\
\includegraphics[angle=-90,scale=.33]{f8g.eps} \hspace{0.8cm}
\includegraphics[angle=-90,scale=.33]{f8h.eps} \\
\end{center}
\caption{
Infrared $K$- (1.9--2.5 $\mu$m) and $L$- (2.8--4.2 $\mu$m) band spectra 
of NGC 2623, Mrk 266, and Arp 193.
The abscissa is the observed wavelength in $\mu$m, and the ordinate is
flux F$_{\lambda}$ in 10$^{-15}$ W m$^{-2}$ $\mu$m$^{-1}$.
Strong hydrogen recombination lines, Pa$\alpha$ (1.87 $\mu$m), 
Br$\gamma$ (2.17 $\mu$m), Pf$\gamma$ (3.74 $\mu$m), and Br$\alpha$ (4.05
$\mu$m), the molecular hydrogen H$_{2}$(1--0) S(1) line (2.12 $\mu$m),
and the 3.3 $\mu$m PAH emission feature are indicated. 
The mark ``?'' indicates that detection is unclear.
The dashed lines are the adopted continuum levels for the 3.3-$\mu$m PAH
emission feature, and the dotted lines indicate the fittings of the
3.3-$\mu$m PAH emission using the template profile ($\S$4.2.3).
}
\end{figure}

\begin{figure}
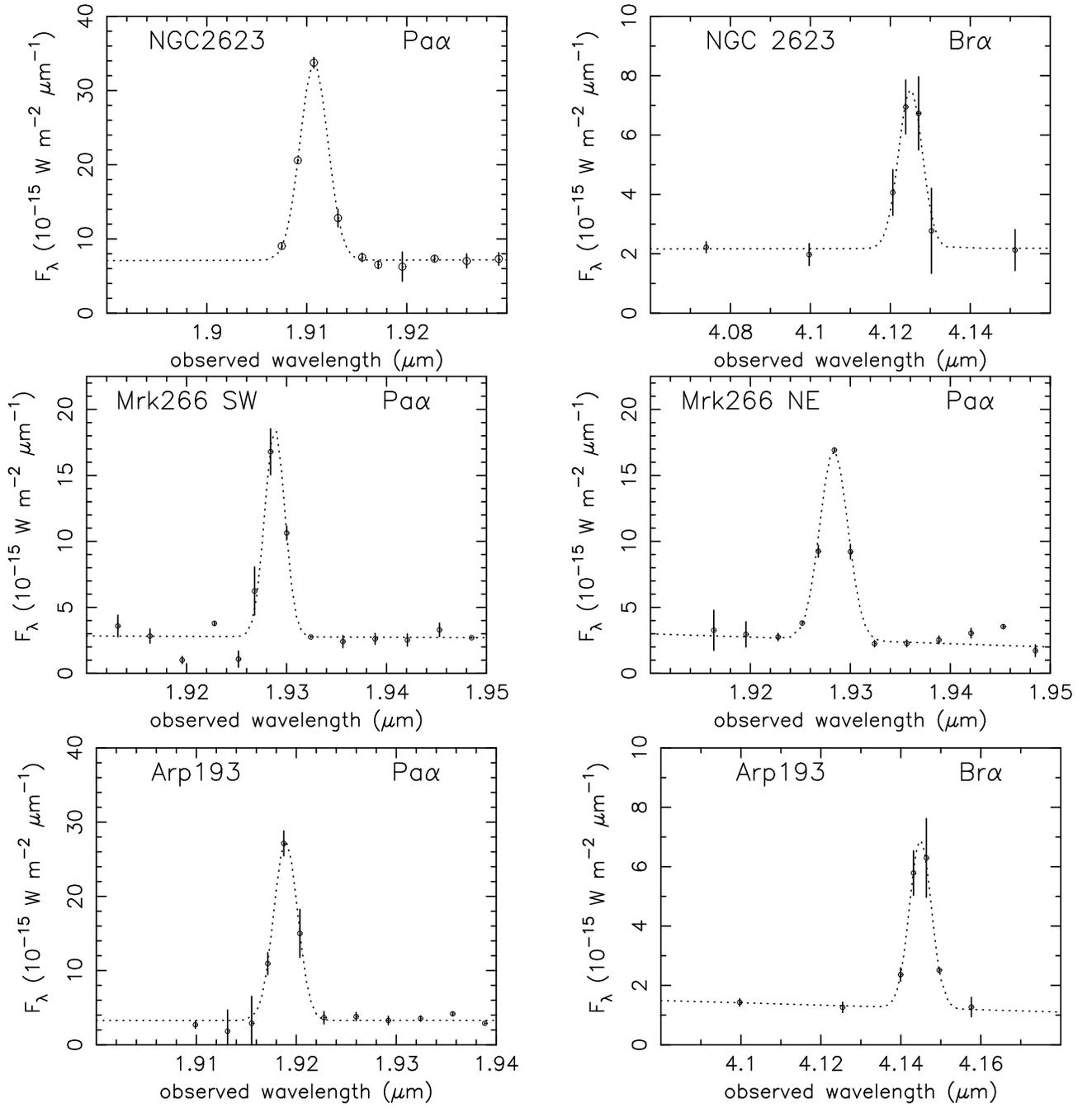

\begin{center}
\includegraphics[angle=-90,scale=.345]{f9a.eps} \hspace{0.8cm}
\includegraphics[angle=-90,scale=.345]{f9b.eps} \\
\includegraphics[angle=-90,scale=.345]{f9c.eps} \hspace{0.8cm}
\includegraphics[angle=-90,scale=.345]{f9d.eps} \\
\includegraphics[angle=-90,scale=.345]{f9e.eps} \hspace{0.8cm}
\includegraphics[angle=-90,scale=.345]{f9f.eps} \\
\end{center}
\caption{
Enlarged spectra around Pa$\alpha$ and Br$\alpha$ emission lines.
The abscissa is the observed wavelength in $\mu$m, and the ordinate is
flux F$_{\lambda}$ in 10$^{-15}$ W m$^{-2}$ $\mu$m$^{-1}$.
Gaussian fits are overplotted as dotted lines. 
}
\end{figure}

\begin{figure}
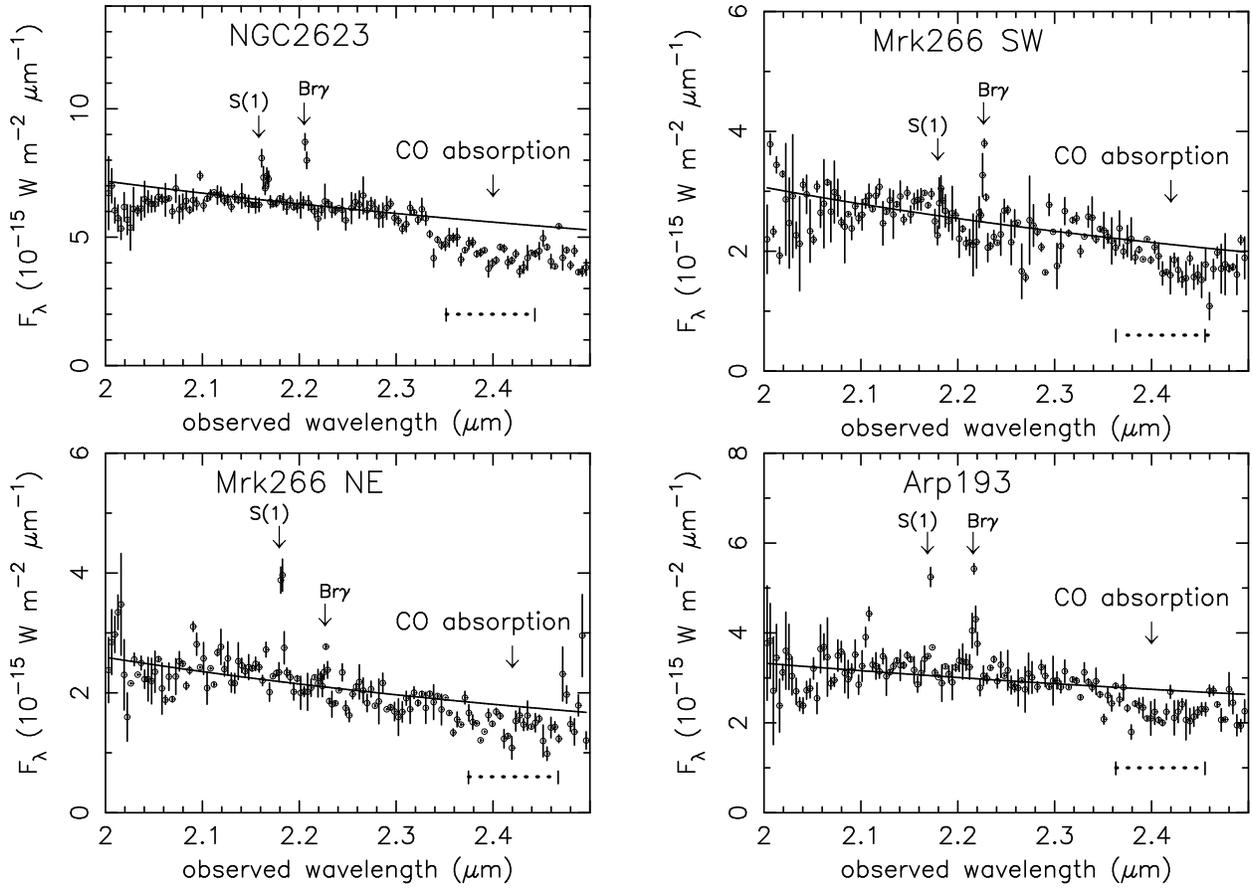

\begin{center}
\includegraphics[angle=-90,scale=.345]{f10a.eps} \hspace{0.8cm}
\includegraphics[angle=-90,scale=.345]{f10b.eps} \\
\includegraphics[angle=-90,scale=.345]{f10c.eps} \hspace{0.8cm}
\includegraphics[angle=-90,scale=.345]{f10d.eps} \\
\end{center}
\caption{
Enlarged spectra around the 2.3--2.4 $\mu$m CO absorption features.
The abscissa is the observed wavelength in $\mu$m, and the ordinate is
flux F$_{\lambda}$ in 10$^{-15}$ W m$^{-2}$ $\mu$m$^{-1}$.
The dotted lines inserted with vertical solid lines mark the wavelength range
used to measure the CO$_{\rm spec}$ values, against the 
adopted continuum levels shown as solid lines.
}
\end{figure}

\begin{figure}
\begin{center}
\includegraphics[angle=-90,scale=.8]{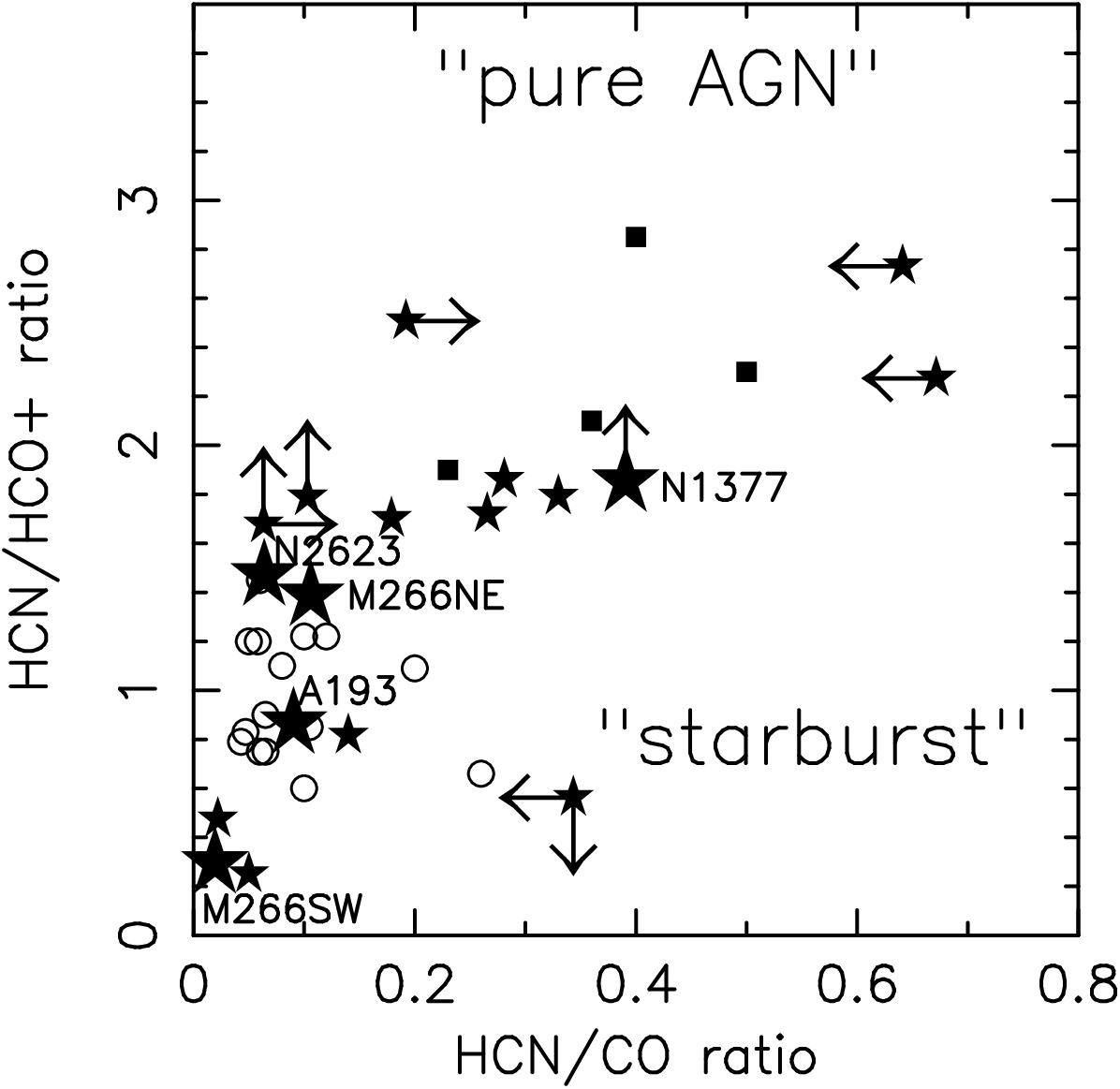} \\
\end{center}
\caption{\small 
HCN(1--0)/HCO$^{+}$(1--0) (ordinate) and HCN(1--0)/CO(1--0) (abscissa)
ratios in brightness temperature ($\propto$ $\lambda^{2}$ $\times$ flux
density), derived from our NMA interferometric observations.   
NGC 2623, Mrk 266 SW and NE, Arp 193, and NGC 1377 are plotted as large
filled stars with labels.   
Other LIRGs previously observed by \citet{ima04,ink06,ima07b} and
\citet{in06} are also plotted as small filled stars.  
Other data points are taken from \citet{koh05}, where sources with
AGN-like (starburst-like) ratios are marked with filled squares (open
circles).
For all LIRG nuclei, the HCN(1--0)/HCO$^{+}$(1--0)
brightness-temperature ratios in the ordinate are those toward the 
nuclei, where putative buried AGNs are expected to reside. 
Contamination from extended star-forming emission outside the beam sizes 
(Table 3) is totally removed.
}
\end{figure}

\begin{figure}
\begin{center}
\includegraphics[angle=-90,scale=.8]{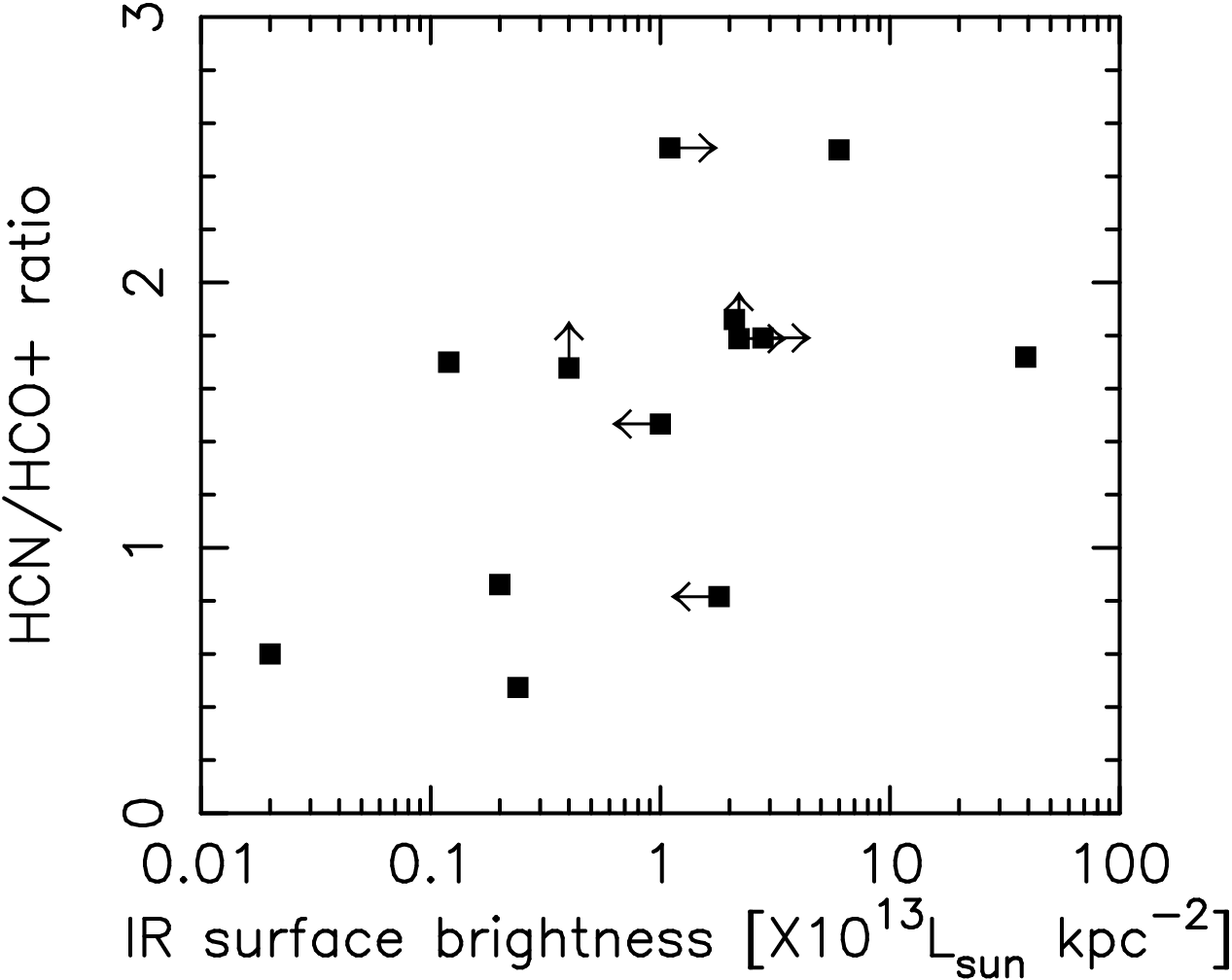}
\end{center}
\caption{
HCN(1--0)/HCO$^{+}$(1--0) brightness-temperature ratio (ordinate) and
infrared emission surface brightness (abscissa) for LIRGs with available
infrared emission surface brightness data \citep{soi00,soi01,eva03}. 
Only LIRGs whose HCN(1--0) and HCO$^{+}$(1--0) lines were simultaneously 
observed with NMA, and the prototypical starburst galaxy M82
\citep{koh01}, are plotted.   
The LIRG sample includes IRAS 08572+3915, UGC 5101, Mrk 231, Mrk 273, Arp
220, IRAS 17208$-$0014, VV 114 E, Arp 193, NGC 2623, Arp 299 A and C, 
and NGC 4418 \citep{ima04,ink06,in06,ima07b}.
For Arp 220, the HCN(1--0)/HCO$^{+}$(1--0) brightness-temperature ratio
is the average value of the eastern and western nuclei \citep{ima07b}. 
}
\end{figure}

\end{document}